\documentclass[aps,pre,twocolumn,showpacs,floatfix,superscriptaddress]{revtex4}
\usepackage{graphicx}
\usepackage{amsmath}
\usepackage{amssymb}

\begin{document} 

\title{Stochastic dynamics of model proteins on a directed graph} 

\author{Lorenzo Bongini}
\email{bongini@fi.infn.it}
\affiliation{Dipartimento di Fisica, Universit\`a di Firenze,
via Sansone, 1 - I-50019 Sesto Fiorentino, Italy}
\affiliation{Centro Interdipartimentale per lo Studio delle Dinamiche
Complesse, via Sansone, 1 - I-50019 Sesto Fiorentino, Italy}
\author{Lapo Casetti}
\email{casettii@fi.infn.it}
\affiliation{Dipartimento di Fisica, Universit\`a di Firenze,
via Sansone, 1 - I-50019 Sesto Fiorentino, Italy}
\affiliation{Centro Interdipartimentale per lo Studio delle Dinamiche
Complesse, via Sansone, 1 - I-50019 Sesto Fiorentino, Italy}
\affiliation{INFN, sezione di
Firenze, via G. Sansone 1, 50019 Sesto Fiorentino, Italy}
\author{Roberto Livi}
\email{livi@fi.infn.it}
\affiliation{Dipartimento di Fisica, Universit\`a di Firenze,
via Sansone, 1 - I-50019 Sesto Fiorentino, Italy}
\affiliation{Centro Interdipartimentale per lo Studio delle Dinamiche
Complesse, via Sansone, 1 - I-50019 Sesto Fiorentino, Italy}
\affiliation{INFN, sezione di
Firenze, via G. Sansone 1, 50019 Sesto Fiorentino, Italy}
\author{Antonio Politi}
\email{antonio.politi@fi.isc.cnr.it}
\affiliation{Istituto dei Sistemi Complessi, CNR, Via Madonna
del Piano, 10 50019 Sesto Fiorentino, Italy}
\affiliation{Centro Interdipartimentale per lo Studio delle Dinamiche
Complesse, via Sansone, 1 - I-50019 Sesto Fiorentino, Italy}
\author{Alessandro Torcini}
\email{alessandro.torcini@fi.isc.cnr.it}
\affiliation{Istituto dei Sistemi Complessi, CNR, Via Madonna
del Piano, 10 50019 Sesto Fiorentino, Italy}
\affiliation{Centro Interdipartimentale per lo Studio delle Dinamiche
Complesse, via Sansone, 1 - I-50019 Sesto Fiorentino, Italy}
\affiliation{INFN, sezione di
Firenze, via G. Sansone 1, 50019 Sesto Fiorentino, Italy}


\begin{abstract}
A method for reconstructing the energy landscape of simple polypeptidic
chains is described. We show that we can construct an equivalent representation
of the energy landscape by a suitable directed graph. Its topological and
dynamical features are shown to yield an effective estimate of the time
scales associated with the folding and with the equilibration processes.
This conclusion is drawn by comparing molecular dynamics simulations
at constant temperature with the dynamics on the graph, defined by a 
temperature dependent Markov process. The main advantage of the graph 
representation is that its dynamics can be naturally renormalized by
collecting nodes into "hubs", while redefining their connectivity.
We show that both topological and dynamical properties are preserved
by the renormalization procedure. Moreover, we obtain clear indications that
the heteropolymers exhibit common topological properties, at variance
with the homopolymer, whose peculiar graph structure stems from its
spatial homogeneity. In order to obtain a clear distinction between
a ''fast folder" and a ''slow folder" in the heteropolymers one has to
look at kinetic features of the directed graph. We find that the
average time needed to the fast folder for reaching its native configuration
is two orders of magnitude smaller than its equilibration time, while for
the bad folder these time scales are comparable.
Accordingly, we can conclude that the strategy described in this paper
can be successfully applied also to more realistic models, by studying
their renormalized dynamics on the directed graph, rather than performing
lengthy molecular dynamics simulations.

\end{abstract} 


\maketitle

\section{Introduction}
\label{secuno}

Numerical simulations are quite often an effective approach for
studying the dynamical properties of systems with many 
degrees of freedom. When the interaction among the constituent particles 
(atoms, molecules, monomers, etc.) are ruled by short--range
forces, molecular dynamics (MD) techniques provide 
useful hints for understanding complex dynamical phenomena.
The main practical limitation to this basic approach is the existence of
relaxation processes, whose time scales are several orders of magnitude
longer than the typical time scale of the microscopic dynamics. Models of 
structural glasses exhibit such puzzling features, which are associated with 
the anomalously high viscosity of these amorphous materials and with the 
ageing phenomenon \cite{refuno}. Something
similar occurs in proteins, where the folding or the equilibration
processes may proceed over exceedingly long time scales with respect to
the microscopic ones \cite{refdue}. In all of these cases reproducing the 
interesting phenomena from a microscopic description can be very expensive in 
terms of CPU time and mass storage memory.
A possible solution to such difficulties could be based on a drastic
simplification of the microscopic model to a "coarse grained"  version,
yielding an effective representation of the dynamics in terms of macroscopic
modes. For instance, such a strategy is usually applied for obtaining a 
hydrodynamic description, by separating the corresponding long time
scales from the fast ones. The so--called "projection technique" at the
basis of the linear--response theory by Green and Kubo \cite{GK}
provides a suitable tool for cooking such a recipe. Unfortunately, it 
applies successfully only to a few simple models of hard--sphere fluids.

A suitable strategy for tackling an effective description of anomalously 
long relaxation processes has to take into account that they are  
associated with the peculiar structure of the energy landscape.
In particular, the phase space is typically partitioned hierarchically
into loosely connected regions: a trajectory may last over a very long time
inside a subset of the available phase space, before finding a wayout
through some "bottleneck" and enter an unexplored region. For 
temperatures small enough with respect to the energy barriers
of the bottlenecks, the situation may look quite similar to the phenomenon
of the breaking of ergodicity 
characterizing the phase transitions of statistical mechanics. On the
other hand, since in MD the number of degrees of freedom
is finite and the phenomenological potentials are typically represented
by smooth functions, one has not to deal with real singularities and we can
guess that the entire phase space should be explored after a sufficiently
long (usually, exceedingly long) time. It is worth stressing that this 
conjecture seems quite plausible, despite no general rigorous proof of 
ergodicity for such phenomenological potentials is available.

On a finer spatial scale one can view the energy landscape as a
collection of the basins of attraction of the local minima of the
potential energy.  A basin of attraction of a local minimum is the collection 
of points in phase space, whose overdamped dynamics converges to that minimum.
By associating a different symbol to each basin of attraction one could reconstruct
a dynamical trajectory as a sequence of such symbols, thus yielding a coarse--grained
representation of the dynamics. At the boundaries between nearby basins of attractions
there are saddles which allow a trajectory to explore different basins during its
evolution.  We guess that a suitable approximation of the dynamics in phase 
space can be obtained by replacing the trajectories with a sequence of thermally
activated transitions between different local minima. 
We expect that such an approximation is effective when the temperature of the 
system (expressed in units of the Boltzmann constant) is small enough to be compared with
the typical height of the energy barriers (saddles) separating nearby basins.

Relying upon these considerations, we assume that for sufficiently (but no too)
small temperatures the MD can be effectively replaced by a 
stochastic dynamics defined onto a connected graph. The local minima, or
equivalently their basins of attraction, are the
nodes of this graph. A link between two nodes is drawn if the corresponding minima 
are connected in phase space by a saddle. The strength associated with the link is
given by the transition rate between the two coupled minima. This quantity can be
determined by purely geometric features of the energy
landscape as a suitable generalization of the Arrhenius law to a high--dimensional
space (i.e., Langer's formula, see Section \ref{due3}).
The directed nature of the graph is a straightforward consequence of the different
energy gaps which, in general, separate each one of the two minima from their connecting 
saddle.

We want to stress that the equivalence between the MD and the
stochastic coarse-grained dynamics on the directed graph has to be assumed valid
in a statistical sense. Actually, a stochastic path through the graph should not necessarily 
correspond to any dynamical trajectory. Nonetheless, we guess that
by averaging over many paths on the directed graph one obtains 
statistical inferences consistent with averages over many trajectories generated by
the MD.

In this manuscript we test this approach by studying simple model proteins
whose dynamical features were analyzed in a previous publication \cite{mc}. 
In Section \ref{secdue} we summarize the model and the  method used for reconstructing minima
and saddles of the energy landscape of such model proteins. We also show that 
the strategy proposed in \cite{mc} can be improved by adopting a suitable
search procedure for identifying shortcuts, which connect minima separated by
large conformational distances. Later in this section we show how one can
plug a Markov--chain structure onto the reconstructed directed graph.
In Section \ref{sectre} we discuss how such a representation can be used for extracting equilibrium
and non equilibrium properties, to be compared with MD simulations.
As expected the "graph" approximation is effective in a range of temperatures
close to the folding one. In Section \ref{secquat} we comment about the advantages of
the "graph" approximation with respect to MD simulations. In fact, one can easily
realize that a suitable reconstruction of the energy landscape,
including a sufficient sampling of minima and saddles, requires a considerable 
numerical effort, comparable with MD simulations. This notwithstanding,
a graph representation allows to obtain many additional information on the
main features of the model. In particular, we argue that a renormalization
procedure can be successfully applied in such a way to keep the main dynamical
and statistical features associated with equilibrium and transient processes,
while eliminating the unessential details associated with a large number 
of minima and saddles. Moreover, additional informations about the nature of
model proteins can be obtained by applying standard analysis of static properties
of random directed graphs. We can conclude that some general static features are
common to any model of a polypeptidic chain. The main specific signatures identifying
a protein specimen ({\sl fast folder}) seem rather associated to dynamical features.
This is not completely unexpected, although quite often one can find in the 
literature claims about specific static properties of the energy landscape as intrinsic
to real proteins \cite{zscore}. Our analysis at least challenges this widespread belief.

\section{The model, its energy landscape and the directed graph}
\label{secdue}

\subsection{A simple toy-model of polypeptidic chains in 2D}
\label{due1}
 
For the sake of simplicity we want to test our idea of approximating the thermalized 
dynamics of a polypeptidic chain by a stochastic dynamics on a directed graph 
with a toy model, first introduced in \cite{tlp}. This is a slight modification of 
the 2d off--lattice HP model originally proposed by Stillinger {\it et al.} in 
\cite{Still}.
It is defined by the Hamiltonian
\begin{equation}
H = T + V
\label{hamil}
\end{equation}
where 
\begin{equation}
T = \sum_{i=1}^L \frac{p_{x,i}^2+p_{y,i}^2}{2} 
\end{equation}
is the kinetic energy and 
\begin{eqnarray}
V &=& \sum_{i=1}^{L-1} V_1(r_{i,i+1}) + \nonumber\\
&+&\sum_{i=2}^{L-1} V_2(\theta_i)+ \sum_{i=1}^{L-2} \sum_{j=i+2}^{L}  V_3(r_{ij},\xi_i,\xi_j)
\end{eqnarray}
is the potential energy defined by the phenomenological potentials
\begin{eqnarray}
V_1 (r_{i,i+1}) &=& \alpha (r_{i,{i+1}}-r_0)^2, \nonumber\\
V_2(\theta_i) &=& \frac{1 - \cos\theta_i}{16},\nonumber\\
V_3(r_{i,j}) &=& \frac{1}{r_{i,j}^{12}} - \frac{c_{i,j}}{r_{i,j}^6}
\label{v1}
\end{eqnarray}
All the parameters are expressed in terms of adimensional arbitrary units:
for instance, $\alpha$ and $r_0$ are fixed to the values 20 and 1, respectively.
The model Hamiltonian represents a one--dimensional chain of $L$ point--like monomers
corresponding to the residues of a real protein. Only two types of residues are
considered: hydrophobic, H, and polar, P~. Accordingly, a heteropolymer is
identified by a sequence of discrete variables $\{ \xi_i \}$ (with
$i=1, \dots, L$) along the chain: $\xi_i = \pm 1$ 
indicates that the $i$-th residue is of type H or P, respectively. 
The intramolecular potential is composed of
three terms: a stiff nearest--neighbor harmonic potential $V_1$, which keeps
the bond distance almost constant, a three--body potential $V_2$, which measures
the energetic cost of local bending, and a long--range
Lennard--Jones potential $V_3$ acting between all pairs of monomers $i$ and $j$
such that $|i-j| >1$.
The monomers are assumed to have the same unitary mass. 
The space coordinates of the $i$-th monomer are ${\mathbf q}_i=(x_i, y_i)$ and their conjugated 
momenta are ${\mathbf p}_i=(p_{x,i},p_{y,i}) = ({\dot x}_i,{\dot y}_i)$~.
The variable $r_{i,j} = \sqrt{(x_i-x_j)^2+(y_i-y_j)^2}$ is the distance between 
$i$-th and $j$-th monomer and $\theta_i$ is the bond angle at the $i$-th monomer. 
$V_3$ is the only contribution that depends on the nature of the monomers.
The coefficients $c_{i,j} = \frac{1}{8} (1+\xi_i + \xi_j +5 \xi_i \xi_j)$
are defined in such a way that the interaction is attractive if both residues are either
hydrophobic or polar (with $c_{i,j} = 1$ and $1/2$, respectively),
while it is repulsive if the residues belong to different species
($c_{ij} = -1/2$).

Here, we focus our investigation on three sequences
of twenty monomers that represent the three classes 
of different folding behaviors observed in this model:
\begin{itemize}
\item{[S0]} a homopolymer composed of 20 H residues;
\item{[S1]=[HHHP HHHP HHHP PHHP PHHH]} a sequence that has been
identified as a fast--folder in \cite{irback2};
\item{[S2]=[PPPH HPHH HHHH HHHP HHPH]} a randomly generated sequence,
that has been identified as a slow--folder in \cite{tlp}.
\end{itemize}

The three characteristic temperature $T_{\theta}$, $T_f$, $T_g$ of each 
sequence determined in \cite{tap} by MD simulations, 
where the chains are in contact with a Langevin heat reservoir,
are reported in Table \ref{tabone}.
\begin{table}[ht]
\begin{tabular}{|c|c|c|c|c|c|}
\hline
 \hfil & \hfil S0 \hfil & \hfil S1 \hfil &
\hfil S2 \hfil \hfil \\
\hline\hline
$T_\theta$ & 0.16  & 0.11 & 0.13 \\
$T_f$ & 0.044  & 0.061 & 0.044 \\
$T_g$ & 0.022  & 0.048 & 0.025 \\
$n_0$ & 31 & 37 &  36 \\
$V(0)$ & -7.04 & -4.67 & -4.67 \\
\hline
\end{tabular}
\caption{
The collapse transition temperature $T_{\theta}$, the folding temperature $T_f$,
the "glassy" temperature
$T_g$, the number $n_0$ of nearest neighbor
minima of the native configuration for the sequences S0 (homopolymer), S1 (fast--folder) and
S2 (slow--folder). $V(0)$ is the minimum of the potential energy, corresponding to the
native state.}
\label{tabone}
\end{table}

Several distances can be defined in order to distinguish between two configurations ${\cal
C}_1$ and  ${\cal C}_2$ of a two-dimensional chain. A particularly simple one is the angular
distance 
\begin{equation}
\label{angulardistance}
d_{\theta}({\cal C}_1,{\cal C}_2)={1\over L-2} \sum_{n=1}^{L-2} |\theta_i({\cal C}_1)-\theta_i({\cal C}_2)|,
\end{equation}
where
\begin{equation}
\label{angle}
\theta_i ({\cal C})={({\mathbf q}_i-{\mathbf q}_{i+1}).({\mathbf q}_{i+1}-
{\mathbf q}_{i+2})\over r_{i,i+1}r_{i+1,i+2}}
\end{equation}
is the $i$-th backbone angle of configuration ${\cal C}$, represented by the coordinates
${\mathbf q}_i$.

\subsection{Reconstruction of the energy landscape}
\label{due2}

A suitable reconstruction of the energy landscape
can be obtained by effective computational strategies for identifying 
the local minima of the potential energy and the first order saddles connecting them
in a high--dimensional configuration space. 
A first extensive search for minima can be performed by
sampling a sufficiently large set of MD trajectories 
at a given temperature $T$. This is fixed by a Langevin heat bath 
acting on all the degrees of freedom of Hamiltonian (\ref{hamil}). Since one is 
usually interested in exploring the effects of the energy landscape on the folding
process of polypeptidic chains, it is convenient sampling trajectories at fixed time 
intervals close to the folding temperature $T_f$. Each sample is a dynamical 
configuration, which is used as the initial condition for the overdamped dynamics
(see \cite{mc}): it eventually converges to 
the local minimum, whose basin of attraction contains the sampled configuration.
As shown in \cite{mc} a large number of minima of the energy landscape of sequences
S0, S1 and S2 can be identified by performing {\cal O}($10^3$) Langevin trajectories 
at $T=0.1$ and sampling them at a time pace $\Delta t = 0.1$ in the natural time units of the
model. The strength of the  coupling with the heat bath is given by the dissipation
rate $\gamma = 7$: this value, expressed in the adimensional units of the model, has
been estimated from the knowledge of the relaxation rate of an aminoacid in a
solvent (typically, water) \cite{tap}.

Once this preliminary set of minima has been produced, one is interested in 
determining the pairs of minima which are connected through a first order
saddle in the landscape. 
In fact, for potentials of class ${\mathrm C}^2$ or higher, as those contained in 
Hamiltonian (\ref{hamil}), the minimal energy path connecting two minima,
usually passes through a first order saddle of the energy landscape \cite{fos}.

In the last decade various methods have been proposed
for designing an efficient algorithm for the search of saddles
\cite{wales-eigenvector,mousseau,vanErp}. Unfortunately, none of these statistical 
approaches provides the detailed reconstruction of the energy landscape to allow 
for an effective representation of the corresponding directed graph.
As shown in \cite{mc}, one can take advantage from a metric criterion for identifying 
such pairs of minima: they are typically separated by an angular distance
$d_{\theta}$ smaller than the threshold value $d_{\theta}^{thr}= 0.2 $ (see \cite{mc}).
On the other hand, despite being quite effective, one cannot expect that this criterion 
can identify all the relevant saddles involved in the folding process.
Actually, it has been observed that the energy landscapes of sequences S1 and S2 
contain a relatively small set of first order saddles connecting pairs of minima, 
whose $d_{\theta}$ is definitely larger than $d_{\theta}^{thr}$.
In order to avoid missing such saddles, one can use a more 
refined strategy (see \cite{mc})~. The identified minima are taken as initial conditions 
of the Langevin dynamics at $T=T_f$. The dynamics evolves for a time interval $\Delta \tau =
10^{-3} $ in the natural time units of the model. 
The overdamped dynamics is then applied starting from the final
configuration \footnote{The procedure can be refined and accelerated by applying quasi-Newtonian 
algorithms \cite{qNew}}: if it converges to the starting minimum one goes back to
the final configuration and makes the Langevin dynamics further evolve for a time
interval $\Delta \tau$~. The algorithm is iterated until the overdamped dynamics
converges to a minimum different from the starting one. The new identified minimum is
added to the database of minima, if it had not been previously recorded. Accordingly,
the database of saddles is also updated by adding a connection between the starting minimum
and the new one. This procedure has been repeated ten times from each minimum.


Finally, the first order saddles have been identified by applying the
same procedure proposed in \cite{mc}. Here, we just want to recall that
it is based on an iterative algorithm, which exploits the steepest--descent 
dynamics generated close to the ridge, separating the basins of 
attraction of two connected minima.

The number of minima $N$ and saddles $S$  obtained for each sequence are reported in 
the first two lines of Table \ref{databases}. 
We have also selected the subsets of minima and saddles, whose potential
energy is contained in between the potential energy of the native state, $V_0$
(see Table \ref{databases}) and the ''folding energy" 
$E_f=V_0 + 1/2 k_B T_f (2L -3)$~. We expect that these subsets should contain
the main elements associated with the folding process. Their numbers, indicated 
by $N_f$ and $S_f$, respectively, are also reported in Table \ref{databases}. 
Notice that the fraction of minima and saddles below $E_f$ reduces much more
for the heteropolymer sequences (S1 and S2) than for the homopolymer (S0).
This is a first indication that the energy
landscape close to the native valley exhibits quantitative
differences for different sequences.

\begin{table}
\vskip 0.3 truecm
\begin{tabular}{|l|c|c|c|}
\hline
 \hfil & \hfil S0 \hfil & \hfil S1 \hfil &
\hfil S2 \hfil \hfil \\
\hline\hline
$N$  &  180156 &  87580 &  110524 \\
$S$ &  349197 &  213219 &  304303 \\
$N_f$  & 99797  & 17726 &  35852 \\
$S_f$  &  276958 &  85014 &  150809 \\
\hline
\end{tabular}
\caption{Number of minima, $N$,  and saddles, $S$,  for the three analyzed sequences. 
The number of minima, $N_f$ and saddles, $S_f$, below the folding energy 
$E_f= V_0+1/2 K_B T_f (2L-3)$ is also reported.}
\label{databases}
\end{table}

\subsection{The directed graph}
\label{due3}

A sufficiently rich database containing the minima  and the first order saddles
can be used for constructing  the corresponding directed graph. Each node of
the graph corresponds to one minimum in the database (or, equivalently, to its
basin of attraction). In what follows we  assume that the $N$ nodes are ordered
for increasing values of their potential energy and assign them an index $i$
that runs  from 1 to $N$. Accordingly $i=1$ corresponds to the so--called
native state. A link is traced  in the graph between node $i$ and node $j$ if
the connection database contains a first order saddle $s_{i,j}$ connecting
them.  We associate to this link its "strength", determined by the rate of 
transition from $i$ to $j$, $\Gamma_{i,j}$. This quantity can be  approximated,
for sufficiently low  temperatures, by the Langer estimate \cite{hanggi}. It
generalizes the usual Arrhenius formula by including the entropic factors
associated with the curvatures of minima and saddles in a high dimensional
space:

\begin{equation}
\Gamma_{i,j} = \frac{\omega_{\parallel\;i,j}}{\pi \gamma}
\frac{\prod_{k=1}^{L^\prime} \omega_{i}^{(k)}} {\prod_{k=1}^{L^\prime-1} 
\omega_{\perp\;i,j}^{(k)}}  \exp  \left( -\frac{V(s_{i,j}) - V(i)}{k_B T} \right)
\quad .
\label{jumprate}
\end{equation}
The $\omega_{i}^{(k)}$ are the $L^\prime=2L-3$ nonzero eigen--frequencies of the
minimum associated with node $i$, the $\omega_{\perp\;i,j}^{(k)}$ are the $L^\prime-1$
nonzero frequencies corresponding to the contracting directions 
of $s_{i,j}$, while $\omega_{\parallel\; i,j}$ is the frequency
associated with the only expanding direction of $s_{i,j}$~. The dissipation
rate $\gamma$ is the same used for defining the Langevin dynamics mentioned in
the previous Section.
The exponential factor in (\ref{jumprate}) depends on the height of the energy barrier 
$V(s_{i,j}) - V(i)$, where we have used the simple notation 
$V(s_{i,j})$ and $V(i)$ for the values of the potential energy at saddle
$s_{i,j}$ and at node $i$, respectively. 
Notice that the Langer estimate relies upon the assumption that at
temperature $T$ the rate of transition between two nodes (i.e. minima of the potential
energy) $i$ and $j$ is determined by the potential energy and by the curvature of the
highest-energy point (i.e., the saddle $s_{i,j}$) along the minimal energy path 
connecting $i$ to $j$. 

We expect that a good estimate of the link strength $\Gamma_{i,j}$
is crucial for obtaining a suitable reconstruction of the folding process. For
instance, it has been shown in \cite{mc} that, for fast--folding heteropolymers
like S1,
conformationally distant minima may be connected by saddles with relatively high
transition rates. This observation nicely fits with the image that the 
wandering of a fast--folder in its energy landscape close or below $T_f$
is significantly biased towards its native state by "shortcuts" connecting
quite different configurations with native-like ones.

We conjecture that MD simulations at temperature $T$ can be
effectively replaced by a Markov process on the directed graph, where in
general $\Gamma_{i,j} \not = \Gamma_{j,i}$.

This directed graph representation of the energy landscape naturally yields
a mathematical description in terms of a nonsymmetric $N\times 
N$ connectivity matrix $\Gamma$, whose nonzero elements $\Gamma_{i,j}$ 
are defined in (\ref{jumprate}). Notice that, by definition, $\Gamma_{i,i} \equiv
0 \,\,\, \forall i ~$.

The master equation which defines the time evolution of the 
probability $P_i(t)$ that the polymer is in node $i$ at time $t$
reads:

\begin{equation}
\frac{{\text d} P_i(t)}{\text dt} = \sum_{j=1}^{N}  P_j(t) \Gamma_{j,i}- P_i(t)
\sum_{j=1}^{N}  \Gamma_{i,j} \,\,\, .
\label{masterequation}
\end{equation}

This master equation can be cast into matrix form: 
\begin{equation}
\frac{{\text d} P(t)}{\text dt} = - W P(t)
\label{matrixmast}
\end{equation}
where $P(t)$ is the vector of dimension $N$ at time $t$, whose elements are
the $P_i(t)$, while the entries of the Laplacian matrix 
$W$ are given by the expression
\begin{equation}
W_{i,j}= \delta_{i,j}\sum_{k=1}^{N}  \Gamma_{j,k}  - \Gamma_{j,i} \,\,\, .
\label{defW}
\end{equation}
$W$ is a nonsymmetric real matrix with positive diagonal elements and whose 
rows and columns sum up to zero: according to Gershgorin's theorem \cite{gersh},
all its eigenvalues $r_i$ , $i=1,\cdots, N$, are real and positive, apart the null 
eigenvalue, $r_1 = 0$ (usually, the $r_i$'s are listed in increasing order with the
index $i$). We denote with $w^{(i)}$ the corresponding eigenvectors. 

A computational gain for the numerical diagonalization of $W$ can be obtained
by transforming $W$ to the symmetric matrix ${\cal W} = T^{-1} W T$ where
\begin{equation}
\label{symmetrization}
T=\left(\begin{array}{ccc}
\sqrt{w^{(1)}_1}    &        &       0 \\
                  & \ddots &          \\
 0                &        &  \sqrt{w^{(1)}_N}    
\end{array}\right).
\end{equation}
Actually, it can be shown \cite{vankampen} that this is a consequence
of the validity of the detailed balance conditions, which hold for dynamics (\ref{matrixmast}). 
In fact, the eigenvector $ w^{(1)}$ represents the stationary probability measure on
the directed graph, since it is the solution of the equation $\dot P = 0$. In particular
the components of $ w^{(1)}$ are given by the following expression:
\begin{equation}
w^{(1)}_i = \alpha \frac{e^{ -\frac{V(i)}{k_B T}} }
{\prod_{k=1}^{L^\prime} \omega_{i}^{(k)}},
\label{stationary}
\end{equation}
where $\alpha$ is a suitable normalization constant, such that $\sum_{i=1}^N w^{(1)}_i = 1$. 
Notice that $w^{(1)}_i$ is the stationary probability for the polymer to be at 
node $i$ and its expression corresponds to the statistical weight of thermodynamic 
equilibrium conditions in the harmonic approximation of minimum (node) $i$.
More generally, by combining (\ref{stationary}) and (\ref{jumprate}) it can be easily
verified that for each pair of connected nodes detailed balance conditions hold:
\begin{equation}
w^{(1)}_i \Gamma_{i,j}  =  w^{(1)}_j \Gamma_{j,i}
\label{detbal}
\end{equation}
so that $ w^{(1)}$ contains all the information concerning
the thermodynamic equilibrium conditions for the directed graph.

The nonzero eigenvalues of $W$, $r_i$, $i=2,\cdots, N$,
represent the relaxation rates to equilibrium of the corresponding eigenvectors
$w^{(i)}$.  Given any initial probability distribution on the directed graph
$P(0) = \sum_{k=1}^N c_k w^{(k)}$ (with $c_k$ real with the normalization 
condition $c_1 =1$, see Appendix \ref{appeA}), the value it will take in $i$ at
time $t$ is given by the expression:
\begin{equation}
\label{decay}
P_i(t)=
\sum_{k=1}^N \;\; c_k\; w^{(k)}_i\;\; e^{-r_k t}
\end{equation}
This expression stems from the orthogonality of the eigenvectors $w^{(i)}$
(see Appendix \ref{appeA}).
One can easily realize that in the limit $t\to\infty$ 
$P_i \to w^{(1)}_i$, i.e. any $P(0)$ will eventually evolve to
thermodynamic equilibrium, as expected.

Another consequence of (\ref{detbal}) is that the stationary
probability flux 
\begin{equation}
\label{probabilityflux}
J_{i,j} = w^{(1)}_i \Gamma_{i,j} = 
\frac{ 
\omega_{\parallel\;i,j}
}{
\prod_{k=1}^{L^\prime-1} \omega_{\perp\;i,j}^{(k)}  
}  
e^{\left( -\frac{V(s_{i,j})}{k_B T} \right)}
\end{equation}
depends only on the energy and on the curvature of $s_{i,j}$.

Let us conclude this section by observing that interesting topological properties of the
directed graph can be studied by introducing the "discrete" connectivity matrix
$\Gamma_\mathrm{d}$, where all the nonzero elements of matrix $\Gamma$ are set to 1.
By replacing $\Gamma_\mathrm{d}$ with $\Gamma$ in (\ref{defW}), one can define
the "discrete" Laplacian matrix $W_\mathrm{d}$.
As we are going to discuss in the
following sections, this matrix contains relevant information about the graph
structure: the power-law behavior of the low-frequency component of its spectral
density determines the spectral dimension of the graph \cite{burioni}.
This extends the concept of Euclidean dimension to graphs which are not
defined on a regular lattice.

\section{Comparison between MD and the Markov chain on the directed graph}
\label{sectre}

In this section we want to check if the stochastic dynamics defined by the
Laplacian matrix $W$ on the directed graph is consistent with MD simulations,
at least for values of the temperature $T$ close to $T_f$.
Despite the many approximations introduced in the procedure for identifying minima 
and saddles and in the estimate of the transition rates $\Gamma_{i,j}$, we
expect that a sufficiently detailed reconstruction of the directed graph can
yield statistical results in quantitative agreement with averages over MD
trajectories.

\begin{figure}[h]
\includegraphics[clip,width=6.5cm]{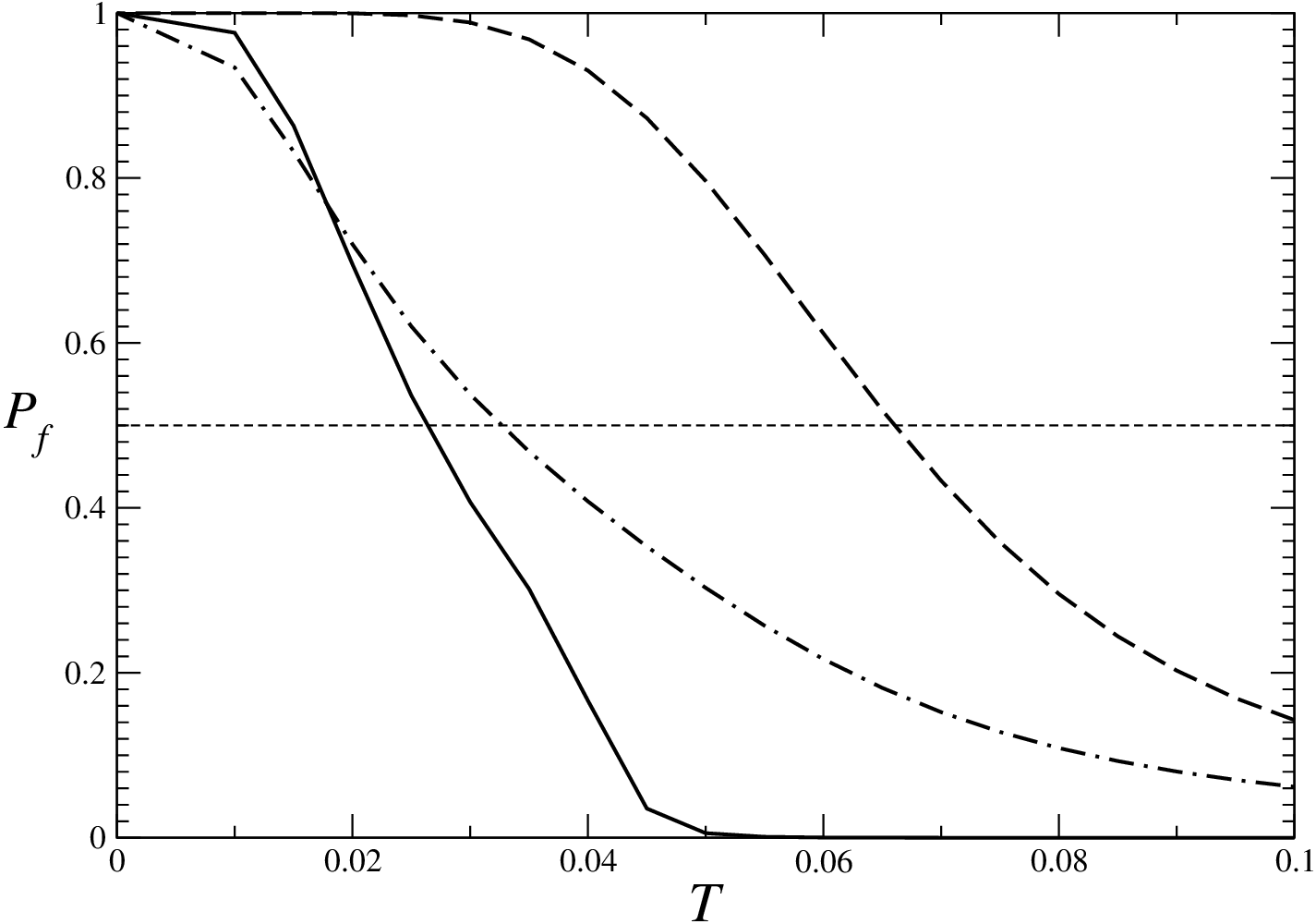}
\caption{Folding probability $P_f$ estimated from Eq.(\ref{stationary})
as a function of temperature $T$.
The full, dot--dashed and dashed lines correspond to S0, S2 and S1,
respectively. }
\label{foldtemp}
\end{figure}

A first simple test can be performed on the expectation values of equilibrium
properties. These can be analytically computed on the graph through the stationary
probability vector $w^{(1)}$  defined in eq.(\ref{stationary}). Since equilibrium properties
depend only on the identified minima, this test can provide 
a quantitative verification of the reliability of the algorithm used for 
locating them in the energy landscape.
In particular, we have compared equilibrium MD estimates of the folding temperatures
$T_f$ of the considered sequences with the same quantities computed by the
equilibrium probability distribution, represented by the eigenvector  $w^{(1)}$. 
In Fig.\ref{foldtemp} we plot the equilibrium probability $P_f$
that a sequence is in the native state or in the set of minima directly connected
with it as a function of temperature $T$. In practice this amounts to measure the 
fraction of ''folded" sequences at a given $T$. We apply the same criterion used in 
equilibrium MD simulations (see \cite{mc}): 
the value of $T_f$ is determined by assuming that at this temperature $50 \%$ of the 
polymer configurations are in the ''folded" state. The results obtained with MD and 
with the directed graph representation are reported in table \ref{tabtemps}.
The quantitative agreement is very good for S1 and
worsens for S0 and S2. Anyway, some 
relevant features are reasonably reproduced by the graph calculation:
both S0 and S2 have similar values of $T_f$, smaller than the values obtained for
S1.

\begin{table}
\vskip 0.3 truecm
\begin{tabular}{|l|c|c|}
\hline
\hfil & \hfil MD \hfil & \hfil graph \hfil  \\
\hline\hline
S0 & 0.044 & 0.026  \\
S1 & 0.061 & 0.066  \\
S2 & 0.044 & 0.033  \\
\hline
\end{tabular}
\caption{folding temperatures as computed by means of MD simulations (table \protect\ref{tabone})
and of the analytical expression of the stationary distribution on the connectivity graph.}
\label{tabtemps}
\end{table}
These results confirm that equilibrium properties of the directed graph
are quantitatively consistent with MD predictions. 

The comparison between MD and the graph dynamics (GD) concerns dynamical indicators.
We report hereafter the results of numerical simulations for the average exit
time from a region in the landscape and for the first passage time from the native state.
The former aims at verifying the conjecture that MD
corresponds to a sequence of thermally activated transitions through the nodes of the
directed graph; the latter amounts to obtain a quantitative estimate of the time
scale involved in the folding process. 

GD is performed by assigning the average time spent at node $i$ as follows 
\footnote{ Alternatively, one could assign the time spent at node $i$ from an exponential
probability distribution with average $\langle t_i \rangle$; we have checked that using this
different recipe yields results similar to those obtained by applying definition
(\ref{jumptime})} :
\begin{equation}
\label{jumptime}
\langle t_i \rangle ={1\over \sum_k \Gamma_{i,k}}.
\end{equation}

The index $k$ runs over all the nodes directly connected with $i$.
The probability of moving from node $i$ to node $j$ is given by the expression
\begin{equation}
\label{jumpprob}
\Pi_{i,j}=\Gamma_{i,j} \langle t_i \rangle
\end{equation}

Accordingly, a trajectory on the directed graph can be represented by an
ordered set of symbols $(i,j,k, \cdots,m,n)$ labeling the sequence of visited 
nodes: its time duration is given by the expression 

\begin{equation}
t=\sum_{\alpha=i}^n \langle t_{\alpha} \rangle \,\,\, .
\end{equation}

Notice that the time scale which establishes the correspondence
between MD and GD is the inverse of the dissipation rate $\gamma$
(see Section \ref{due2} and  eq.(\ref{jumprate}))~.

In Table \ref{tempiEXIT} we report for two temperatures, $T=T_g=0.4$ and $T=T_f=0.6$,
the average exit time for sequence S1 from the native minimum, from the first shell
(i.e., the set of the 66 minima directly connected with the native one) and from the set
of minima ${\mathcal M}$, whose angular distance $d_{\theta}$ from the native state 
is smaller than 0.4. 
The latter set contains 2341 minima, including the first shell.

MD averages have been performed over $10^3$ trajectories starting from each minimum
in the considered set, while GD averaged have been performed
over $10^4$ stochastic paths. The data exhibits a reasonable agreement, if one considers
that Langer's formula is known to systematically overestimate the actual transition
rates $\Gamma_{i,j}$ \cite{hanggi}~.

Measurements of the first passage time of S1  through its native minimum 
have also been performed by averaging over $10^3$ MD trajectories and over $10^4$ 
GD paths. We have considered three different classes
of initial conditions: the minima in the first shell, the set ${\mathcal M}$,
and the larger set of minima ${\mathcal N}$, whose
potential energy is smaller than $V(0)+ 1/2K_B T_f L' = -3.45$ ( 17726 minima). 
For $T=0.04$ we find a good quantitative agreement between MD and GD
(see Table \ref{tempiARRIVI}), while for $T=0.06$ the first passage time on the
the graph is much smaller then the MD predictions. Apart the approximation due to
the Langer's estimate of the transition rate, the main reason for this discrepancy
has to be attributed to the poor reconstruction of the energy landscape for
high values of the potential energy. In fact, we have checked that by considering the
subset of trajectories which go to the native minimum without escaping from the
set of initial minima, average first passage times converge to much closer values.
This implies that MD visits regions of the energy landscape that are scarcely
accessible to the graph dynamics. Only for temperatures smaller than $T_f$
this effect is highly reduced and MD trajectories tend to remain confined
in a smaller portion of the energy landscape, which is more accurately represented
in the graph.
We want to point out that a finer sampling of the energy landscape would 
demand a sensible computational effort even for producing a small improvement.

Performing similar kinds of measurements for S0 and S2 at $T=0.04$ is practically
unfeasible for MD simulations, since the average exit and folding times increase at 
least of two orders of magnitude with respect to S1. Measurements of the average folding time 
at $T=0.06$ starting from the set of minima with potential energy  below $1/2 K_B T_f L'$ 
give $1.3 \times 10^6$ for MD and $9.8 \times 10^5$ for GD. Since
the quantitative agreement is maintained we can reasonably conjecture that it should
hold also at lower temperatures. Accordingly, we have estimated by GD the average first
passage time of S0 and S2 at $T=0.04$. In
both cases we have found values three orders of magnitude larger than those obtained
for S1. We can conclude that, at variance with MD, the graph dynamics allows to estimate 
the average folding time even for $T < T_f$, thus providing a clear characterization
of a fast folder with respect to generic polymers.


\begin{table}
\vskip 0.3 truecm
\begin{tabular}{|l|c|c|}
\hline
\multicolumn{3}{|c|}{From the native state}\\
\hline
\hfill&  T=0.04 & T=0.06 \\ 
\hline
MD   &  4193 & 164 \\
GD   & 3509 & 85 \\
\hline
\hline
\multicolumn{3}{|c|}{From the 1st shell}\\
\hline
\hfill&  T=0.04 &  T=0.06 \\
\hline
MD    & 23459	& 446	\\
GD    & 13077	& 243	\\
\hline
\hline
\multicolumn{3}{|c|}{From the set ${\mathcal M}$}\\
\hline
\hfill&      T=0.04 &  T=0.06 \\
\hline
MD   & 326855	& 2052	\\
GD   & 107775	& 704   \\
\hline
\end{tabular}
\caption{Comparison of average escape times of S1 from three different sets of initial conditions 
computed by MD and GD simulations at temperatures $T=0.04$ and $T=0.06$. 
The first shell
contains 66 minima, while the set ${\mathcal M}$ contains 2341 minima. }
\label{tempiEXIT}
\end{table}
\begin{table}
\vskip 0.3 truecm
\begin{tabular}{|l|c|c|c|c|}
\hline
\multicolumn{3}{|c|}{From the 1st shell }\\
\hline
\hfill&      T=0.04 & T=0.06\\
\hline
MD    &      5746 & 1153\\
GD    &      4307  & 461\\
\hline
\hline
\multicolumn{3}{|c|}{From the set ${\mathcal M}$}\\
\hline
\hfill&   T=0.04 &  T=0.06 \\
\hline
MD   &     8535  &  4120   \\
GD   &     4434  &  512    \\
\hline
\hline
\multicolumn{3}{|c|}{From the set ${\mathcal N}$ }\\
\hline
\hfill&   T=0.04 & T=0.06  \\
\hline
MD   &     6947  &  5759   \\
GD   &    4566   &  583    \\
\hline
\end{tabular}
\caption{Comparison of average first passage times at the native state from three different 
sets of initial conditions computed by MD and GD simulations 
at temperatures $T=0.04$ and $T=0.06$. 
The first shell
contains 66 minima, the set ${\mathcal M}$ contains 2341 minima and 
the set ${\mathcal N}$ contains 17726 minima
. }
\label{tempiARRIVI}
\end{table}

\section{Renormalization of the directed graph}
\label{secquat}

As we have shown in the previous section MD and GD exhibit a reasonable
quantitative agreement, while numerical simulations of the latter are faster than
the former. On the other hand, a considerable computational price has been payed for
obtaining a suitable reconstruction of the energy landscape. The very advantage of the
directed graph representation stems from the possibility of applying an effective
renormalization procedure. In fact, at a given temperature $T$ the 
directed graph can be renormalized by eliminating all those links (saddles)
corresponding to energy barriers smaller than $K_B T$. In practice, the procedure
can be performed by reducing the pair of nodes connected by such a barrier to
a single node, whose connectivity is renormalized accordingly. More precisely, the
renormalization algorithm goes through the following steps.

\begin{itemize}

\item 
For any pair of connected nodes $i$ and $j$ we compute
\begin{equation}
\label{barrier}
\Delta V_{i,j} = V(s_{i,j}) - V(j)
\end{equation}
if $V(j) > V(i)$;

\item
the $\Delta V_{i,j}$ are listed in increasing order and we choose the
subset 

\begin{equation}
\Sigma = \{ \Delta V_{i,j} : \Delta V_{i,j} \leq K_B T \} \,\,\, ;
\end{equation}

\item 
starting from the smallest element of $\Sigma$ the graph is renormalized by
first assimilating node $j$ with node $i$: 
the renormalized equilibrium eigenvector ${\tilde w}^{(1)}$ and the
renormalized rates $\tilde \Gamma_{k,j}$ are defined as follows:

\begin{equation}
\tilde w^{(1)}_i = w^{(1)}_i + w^{(1)} _j
\end{equation}

\begin{equation}
\tilde \Gamma_{k,i} = \begin{cases}
\Gamma_{k,j} \,\,\, \text{if} \,\,\,  \Gamma_{k,j} = 0 \\
\Gamma_{k,i} +\Gamma_{k,j} \,\,\, \text{otherwise}
\end{cases}
\end{equation}

and

\begin{equation}
\tilde \Gamma_{i,k} = \begin{cases}
\frac{ \Gamma_{j,k} w^{(1)}_j }{ w^{(1)}_j + w^{(1)}_i} \,\,\, \text{if} \,\,\, \Gamma_{k,i} =0 \\
\frac{ \Gamma_{j,k} w^{(1)}_j + \Gamma_{i,k} w^{(1)}_i }{ w^{(1)}_j + w^{(1)}_i} \,\,\, \text{otherwise}
\end{cases}
\end{equation}

\item then we pass to the second lowest value in the set $\Sigma$ and proceed sequentially
until all the elements of this set are renormalized.

\end{itemize}

The renormalized rates  $\tilde \Gamma_{i,j}$ define a renormalized evolution matrix $\tilde W$,
whose equilibrium eigenvector is $\tilde w^{(1)}$. It can be easily argued that detailed balance
conditions (\ref{detbal}) are maintained for the renormalized quantities. Moreover, since it only
involves node removal, the procedure just outlined maintains the original ordering of the graph
nodes.

The renormalization procedure transforms also the discrete connectivity matrix
$\Gamma_{\mathrm d}$ and the discrete Laplacian matrix $W_{\mathrm d}$ 
to $\tilde\Gamma_{\mathrm d}$ and
$\tilde W_{\mathrm d}$, respectively. In the following section we shall first discuss
how the topological properties of the renormalized directed graph depend on
temperature. Afterwards, the corresponding dynamical features will be analyzed.

\subsection{Topological properties of the renormalized connectivity graph}

The renormalization procedure presented in Section \ref{secquat} depends on
the temperature and here we want to analyze how it can change the topological 
properties of the directed graph.
 
\begin{figure}[h]
\includegraphics[clip,width=8.3cm]{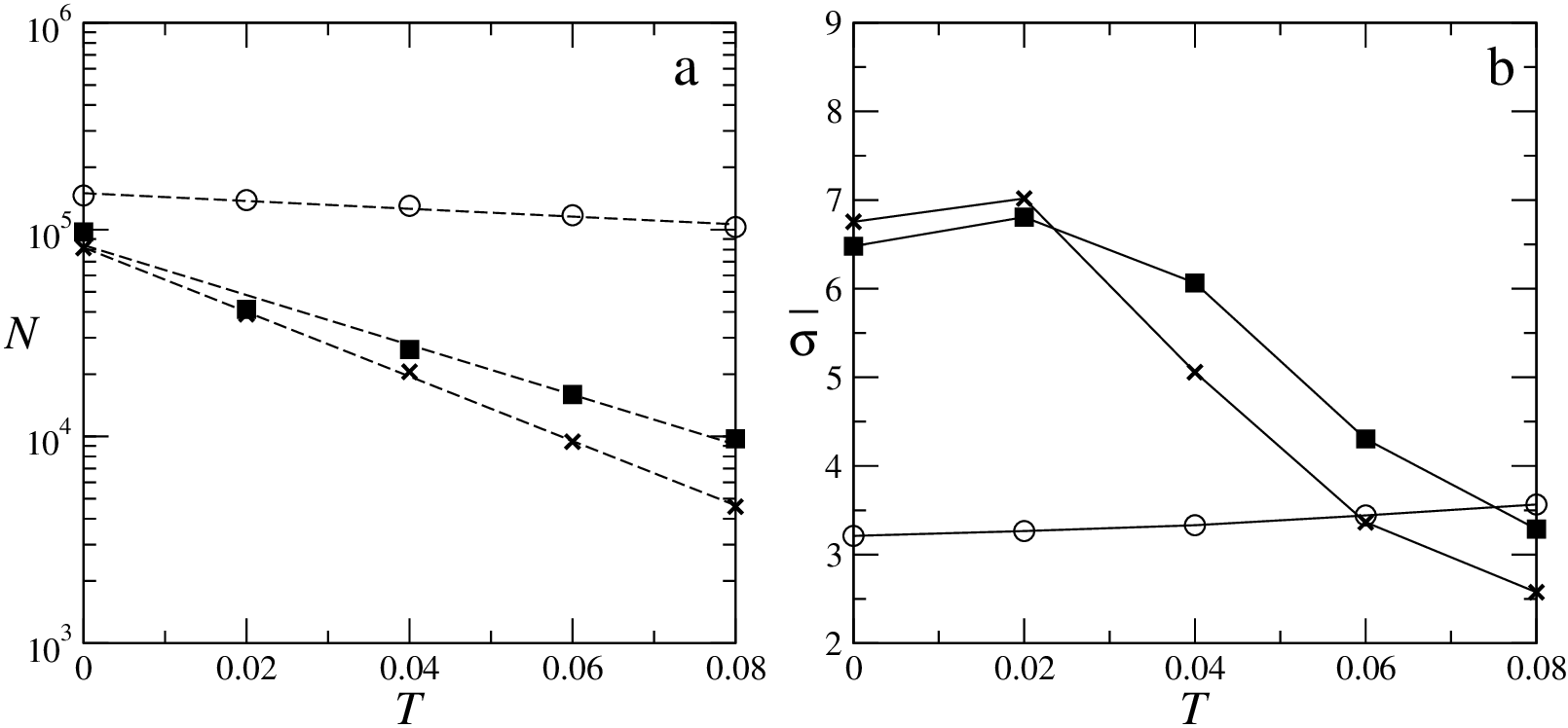}
\caption{Number of nodes $N$ (a) and average connectivity $\bar \sigma$ (b) of the 
renormalized graph  versus temperature $T$ for S0 (empty circles), S1 (crosses)
and S2 (filled squares). In (a) the dashed lines are exponential fits of the data,
due to the adopted log-lin representation. The full lines in (b) are drawn to guide the 
eyes.}
\label{order}
\end{figure}

In Fig. (\ref{order}) we show how  the number of nodes $N$ and the number of
connections per node ${\bar \sigma} = S/N$ change with the temperature, for the 
three sequences 
defined in Section \ref{secdue}. Temperatures are varied in the range $0 \leq T \leq
0.08$, thus encompassing both the glassy and the folding temperatures of all 
sequences (see Table I).
The homopolymer S0 exhibits a very slow exponential decay of $N$, while $\sigma$
slightly increases. Conversely, for S1 and S2, $N$
reduces by more than one order of magnitude, while $\sigma$ uniformly decreases 
beyond $T=0.02 \,$.
These results stress the topological differences between the graph representations of the
homopolymer S0  and of the heteropolymers S1 and S2. In the explored range
of temperatures the graph of S0 is poorly affected by renormalization. This
is due to the peculiar structure of its energy landscape: most barrier heights 
separating local minima from their neighbors are deep enough to keep their role 
of ''hubs" in the renormalized graph, while a relatively small fraction of minima 
are absorbed by the hubs, thus slightly increasing the average connectivity.

\begin{figure}[h]
\includegraphics[clip,width=6.5cm]{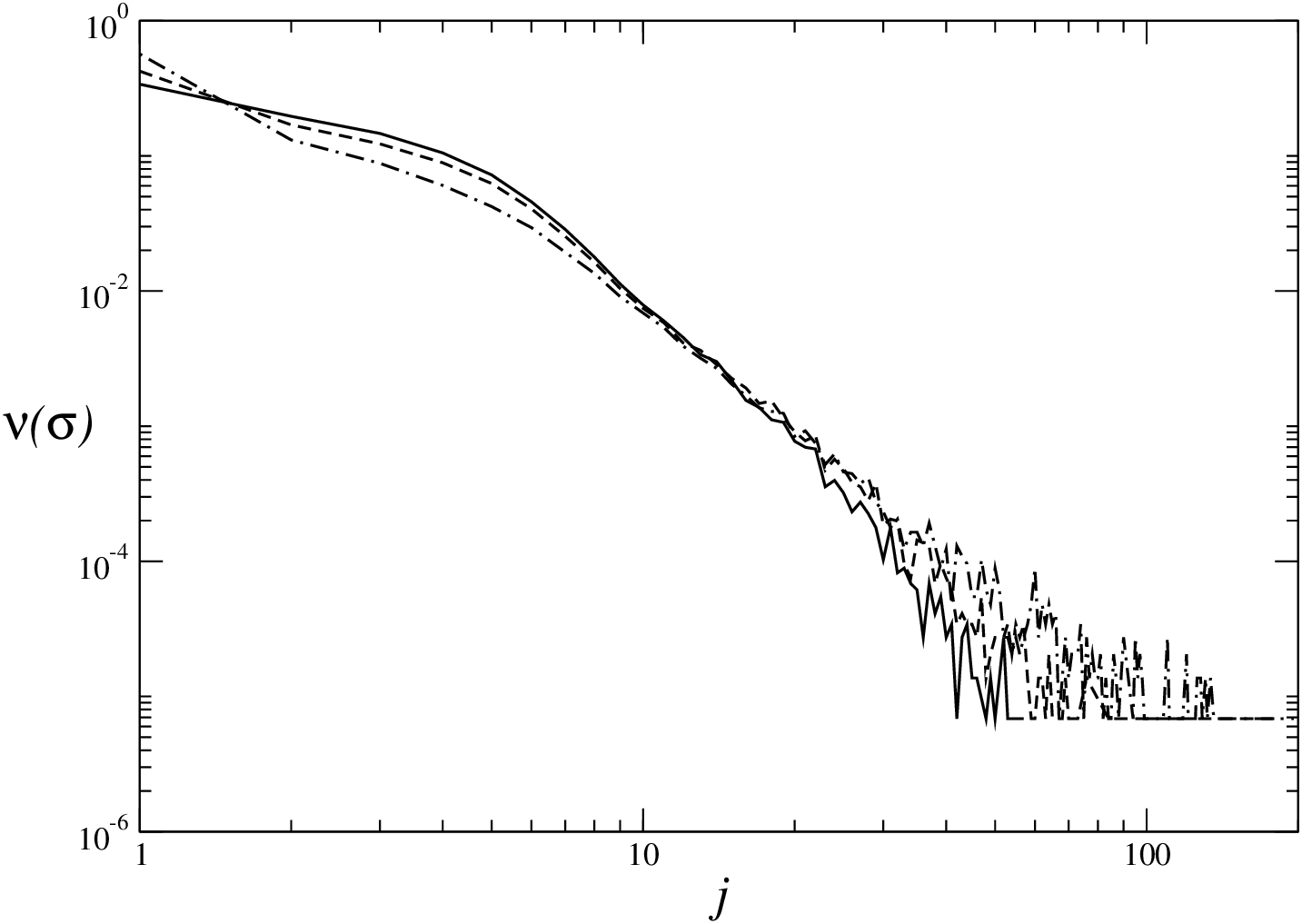} 
\caption{Sequence S0: the fraction of nodes with connectivity $\sigma$, $\nu (\sigma)$,
in log-log scale at $T=0$ (full line), $T=0.04$ (dashed line)
and $T=0.08$ (dot-dashed line).
}
\label{degree00}
\end{figure}

\begin{figure}[h]
\includegraphics[clip,width=6.5cm]{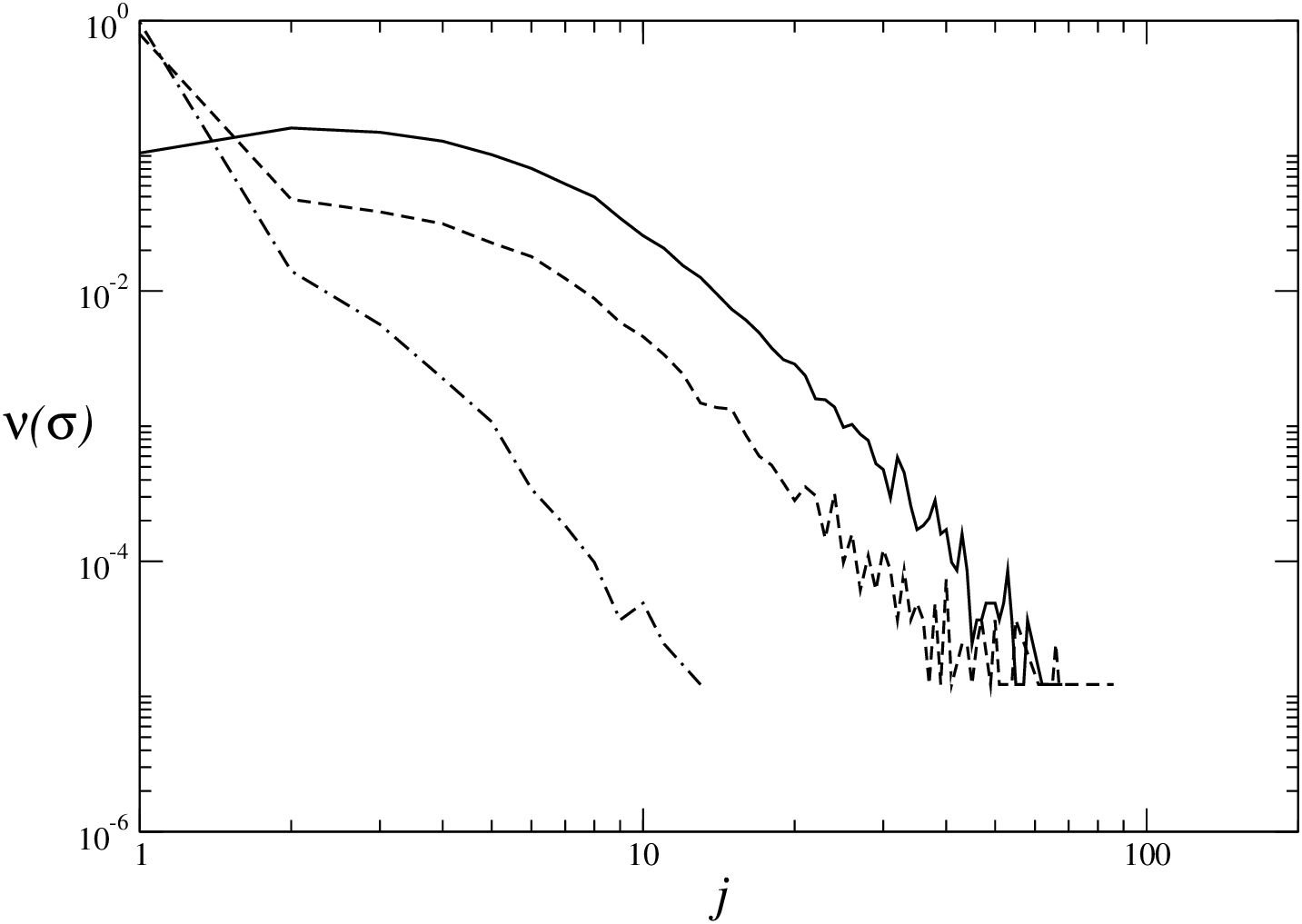}
\caption{
Sequence S1: the fraction of nodes with connectivity $\sigma$, $\nu (\sigma)$,
in log-log scale at $T=0$ (full line), $T=0.04$ (dashed line)
and $T=0.08$ (dot-dashed line).
}
\label{degree81}
\end{figure}

\begin{figure}[h]
\includegraphics[clip,width=6.5cm]{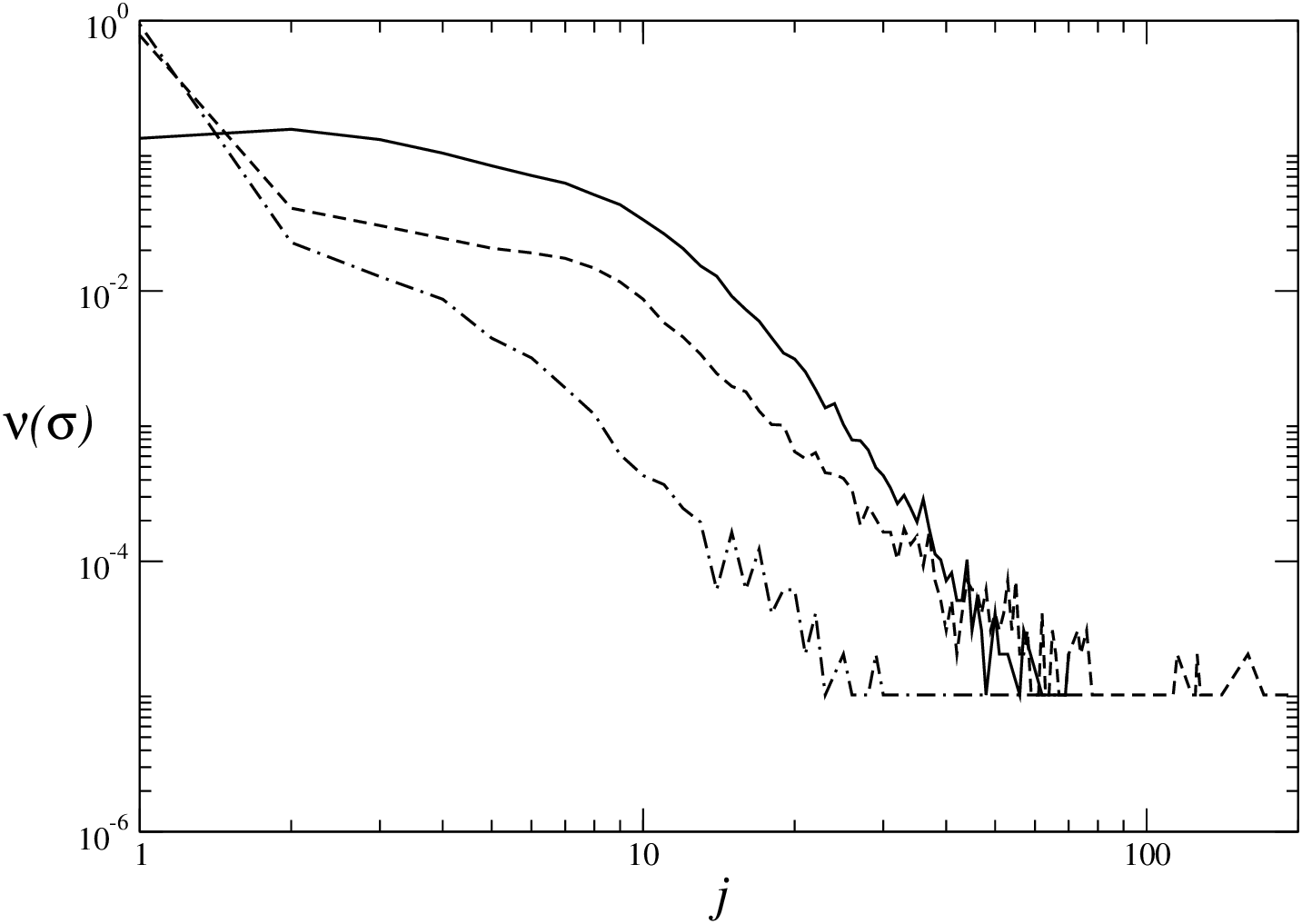}
\caption{
Sequence S2: the fraction of nodes with connectivity $\sigma$, $\nu (\sigma)$,
in log-log scale at $T=0$ (full line), $T=0.04$ (dashed line)
and $T=0.08$ (dot-dashed line).
}
\label{degree01}
\end{figure}

In the graph representation of S1 and S2 the number of hubs is significantly
reduced, because an increasing number of barrier heights is renormalized as temperature 
increases. Moreover, apart an initial increase up to 7, $\sigma$ drops to much
smaller values. The main reason for this sensible reduction of the average connectivity
is essentially due to the fact that most of the hubs share with
the renormalized nodes connections with the same set of neighboring nodes, at variance
with S0, where the connectivity is much more sparse.
A simple argument accounts for these opposite mechanisms. Let us suppose that node
$j$, with connectivity $s_j$, is renormalized to node $i$, with connectivity $s_i$.
The connectivity of the hub $i$ becomes equal to  $
{\tilde s_i}= s_i + s_j - c -2$ if $c$ is the
number of connections with neighboring nodes common to $i$ and $j$. There are two
extreme situations: if $c=0$ (no common connections) then  
${\tilde s}_i= s_i + s_j -2$, while if $c=s_i-1=s_j-1$ ($i$ and $j$ have the
same neighbors) then ${\tilde s_i}= s_i -1$. This argument indicates that above
$T=0.02$ many nodes in the graphs of S1 and S2 are highly connected among
themselves so that renormalization acts as an overall reduction of the average
connectivity. More precisely, only fewer and fewer hubs maintain a high connectivity, 
while the great majority assumes the role of peripheral nodes.
This picture is also confirmed by looking at the distribution of the renormalized average
connectivity $\sigma$ at different temperatures: it is practically unmodified for S0, while
for S1 and S2 it tends to be sharply peaked at 1 as $T$ increases. In Fig.s (\ref{degree00}) ,
(\ref{degree81}) and (\ref{degree01}) we report the fraction of nodes with connectivity $\sigma$,
$ \nu(\sigma) = N(\sigma)/N $, versus $\sigma$ for different values of $T$. Notice that
both $N(\sigma)$ and $N$ vary with the renormalization temperature~.

A special role in the renormalization procedure is played by the 
''native" node, which becomes the main hub of the network, since it collects an 
increasing number of renormalized nodes as $T$ increases. This effect is significantly 
more pronounced for S1 and S2 with respect to S0:
for instance, at $T=0.08$  $\sigma_0 \sim 10^4 $ for S1 and S2, while 
$\sigma_0 \sim 10^3 $ for S0~.

The topological properties of the directed graphs considered in this
section can be also described by determining their spectral dimension $\tilde d$ 
\cite{burioni}. In an infinite graph this quantity is defined by the formula

\begin{equation}
\label{specdim}
R(\omega) \sim \omega^{\tilde d}
\end{equation}
where $R (\omega)$ is the integrated density of harmonic vibrational modes with frequency $\omega$.
It is worth recalling that in our model we are dealing with the spectrum of the discrete Laplacian
matrix $W_\mathrm{d}$, with finite rank $N$ (see Section \ref{due3})~. If we denote its eigenvalues
with $\lambda_1< \lambda_2< \cdots < \lambda_\mathrm{k} < \cdots < \lambda_\mathrm{N}$
we can define the integrated density of eigenvalues $R(\lambda_\mathrm{k})$. In this case, we
can reasonably assume that, if the rank $N$ of $W_{\mathrm d}$ is sufficiently large, by identifying 
$\omega$ with $\omega_k = \lambda_k^{1/2}$ the spectral dimension can be estimated through
the approximate relation

\begin{equation}
\label{specdimapprox}
R(\lambda_k) \sim \lambda_k^{\tilde d/2}
\end{equation}

\begin{figure}[h]
\includegraphics[clip,width=6.5cm]{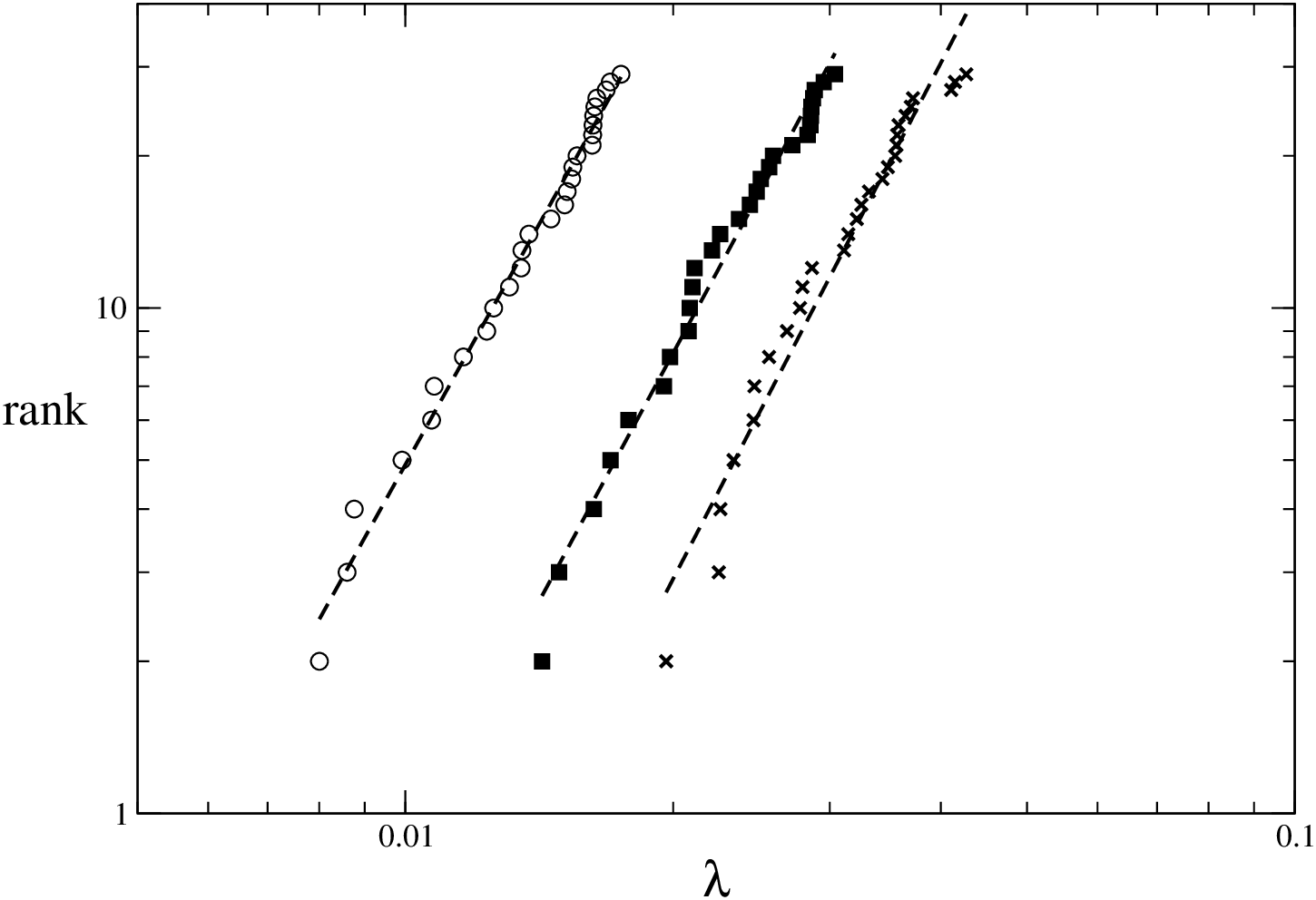}
\caption{Log--log plot of the spectrum of eigenvalues of the discrete Laplacian matrix $W_\mathrm{d}$ of the 
zero--temperature connectivity graph of S0 (empty circles),
S1 (crosses) and S2 (filled squares).
The dashed line refer to power--law with exponents
3.3~ . 
}
\label{spectdimT=0}
\end{figure}

\begin{figure}[h]
\includegraphics[clip,width=6.5cm]{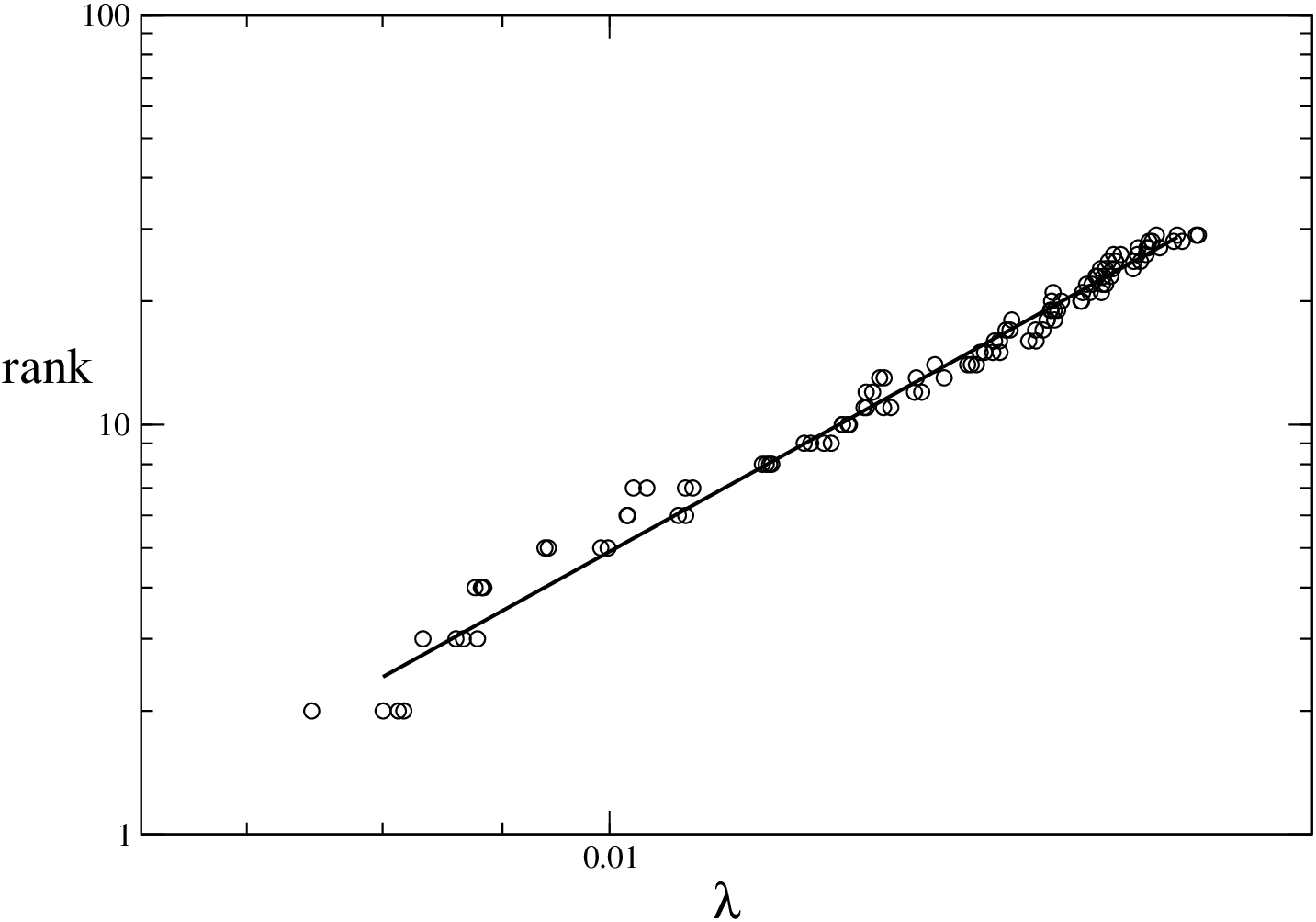}
\caption{Log--log plot of the spectrum of eigenvalues of the discrete Laplacian matrix $W_\mathrm{d}$ 
of S0 for five different temperatures: $T=0.00, 0.02,
0.04, 0.06, 0.08$. We have used the same symbols independently of temperature to
point out that data collapse on the same curve: the full line corresponds to the
power-law fit with exponent 3.1.
.}
\label{spectdim00}
\end{figure}

\begin{figure}[h]
\includegraphics[clip,width=6.5cm]{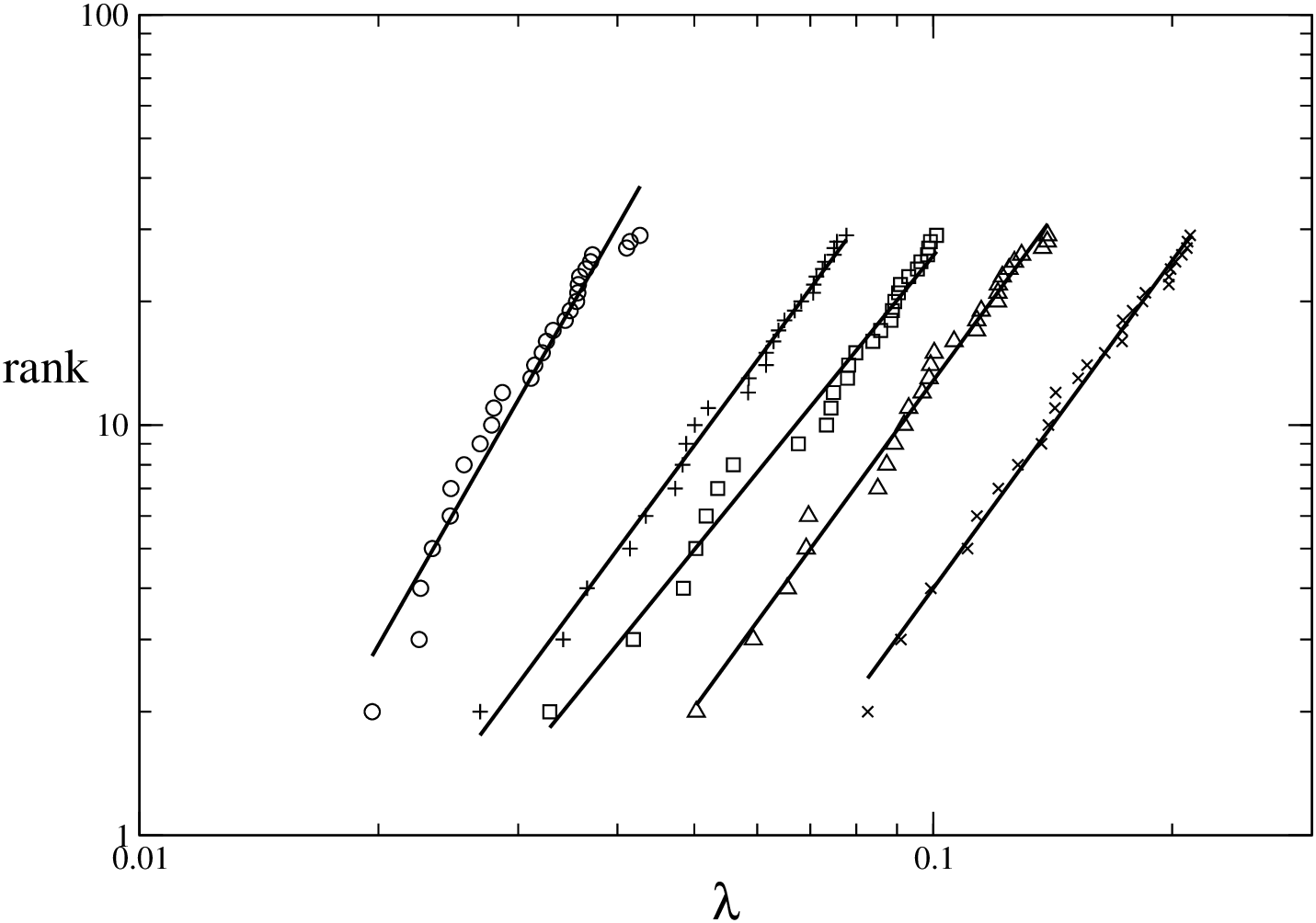}
\caption{Log--log plot of the spectrum of eigenvalues of the discrete Laplacian matrix $W_\mathrm{d}$
of S1 for five different temperatures: $T$=0.00 ({\Large$\circ$}), 0.02 (+),
0.04 ($\Box$), 0.06 ($\triangle$), 0.08 ($\times$). The data obtained for different temperatures are shifted 
horizontally by an arbitrary constant factor in order to obtain a better view.
Notice that the power-law fit passes from its maximum value, 3.4 at $T=0$ to
a minimum value of 2.5~ above $T=0.02$.
}
\label{spectdim81}
\end{figure}

In Fig. \ref{spectdimT=0} we show that one obtains close estimates of the spectral dimension ($\tilde d \approx 6.5$) 
for the connectivity graphs of all
the three sequences considered in this paper. In an infinite graph the application of the
renormalization procedure described in Section \ref{secquat} cannot change the value
of $\tilde d$ \cite{burioni}. This is what happens also for the finite graph representing the homopolymer S0: 
Fig. \ref{spectdim00} shows that, independently of the temperature $T$, the value of $\tilde d$ in the 
renormalized graph is constant. Conversely, for the heteropolymers S1 and S2  the value
of $\tilde d$ decreases to a value close to 5 for $T > 0.02$ (see Fig. \ref{spectdim81}, where we
report the data for sequence S1, which exhibits the same behavior of sequence
for S2). This indicates that, beyond
this temperature, the renormalization procedure modifies the topological properties of the
finite directed graphs: regions of high connectivity collapse into big hubs, while in the renormalized
graph the average connectivity is significantly reduced. This is consistent with what is shown
in Fig. \ref{order}~. 

We can conclude that all topological indicators analyzed in this section allow to distinguish
between the homopolymer S0 and the other heteropolymers. On the other hand, no significant
difference between S1 and S2 can be identified. As we have discussed in Section \ref{sectre}
this seems to emerge only by considering dynamical properties of the directed graphs.
In particular, in what follows we are going to show that this different dynamical
features are maintained also in the renormalized graphs.

\subsection{Relaxation times and the low frequency spectrum of the Laplacian matrix}
\label{seccinq}
One can easily realize that obtaining a complete characterization of the
spectrum of the Laplacian matrix $W$ defined in (\ref{defW}) demands the 
diagonalization of a matrix whose rank is ${\cal O}(10^5)$ (see Table \ref{databases}).
This is practically unfeasible. On the other hand, we are at least interested
in reconstructing the part of the spectrum of $W$ containing the smallest
eigenvalues $r_k$, $k=2,3, \cdots$, which are related to the largest equilibration
time scales. This task can be accomplished by using Lanczos-like algorithms 
of the ARPACK Library \cite{arpack}.

We have first checked that the lowest part of the spectrum
of $W$ and of its renormalized version $\tilde W$ are consistent. In fact,
since the renormalization procedure described in the previous section amounts 
to a series of local transformations on the directed graph, we expect that
the first nonzero eigenvalues should not vary. The corresponding eigenvectors,
$w^{(k)}$, $k=2,3, \cdots$ should as well keep their "structure", modulo the
renormalization procedure, which changes their dimension (see Section
\ref{secquat})
. Such a comparison
has been performed for the directed graph of S1 at $T=0.08$
\footnote {
For the sequences S0 and S2 a direct comparison between
$W$ and $\tilde W$ is practically unfeasible at any temperature, due
to the exceedingly large rank of $W$.}. 
Table \ref{tabranks} reports the rank of $\tilde W$ at various temperatures 
for the three sequences investigated.
\begin{table}[ht]
\begin{tabular}{|c|c|c|c|}
\hline
 \hfil T \hfil& \hfil S0 \hfil & \hfil S1 \hfil &
\hfil S2 \hfil  \\
\hline\hline
0.02 & 138791  & 41139 & 38997 \\
0.04 & 130326  & 20537 & 26292 \\
0.06 & 117328  &  9430 & 15943 \\
0.08 & 102933  &  4576 &  9736 \\
\hline
\end{tabular}
\caption{
Rank of the renormalized Laplacian matrix $\tilde W$ the three investigated sequences at various 
temperatures below $T_\theta$.}
\label{tabranks}
\end{table}

Relying upon this analysis, we can assume that the interesting spectral 
properties of $W$ can be extracted from $\tilde W$ (by exploiting also the
symmetrization procedure described in Section \ref{due3}). In fact, if 
we consider lower values of the temperature, i.e. $T=0.06$ and $T=0.04$,
the rank of $\tilde W$ reduces for S1 to 9430 and 4576, respectively (see Table 
\ref{tabranks}). On the other hand, a direct comparison with the spectrum of 
$W$ would demand a too high computational cost: at lower temperatures the 
relative separation between the eigenvalues at the lower end of the spectrum 
quickly approaches machine precision thus dramatically slowing down the convergence 
rate of the diagonalization algorithm \cite{arpack}.

\begin{figure}[h]
\includegraphics[clip,width=8cm]{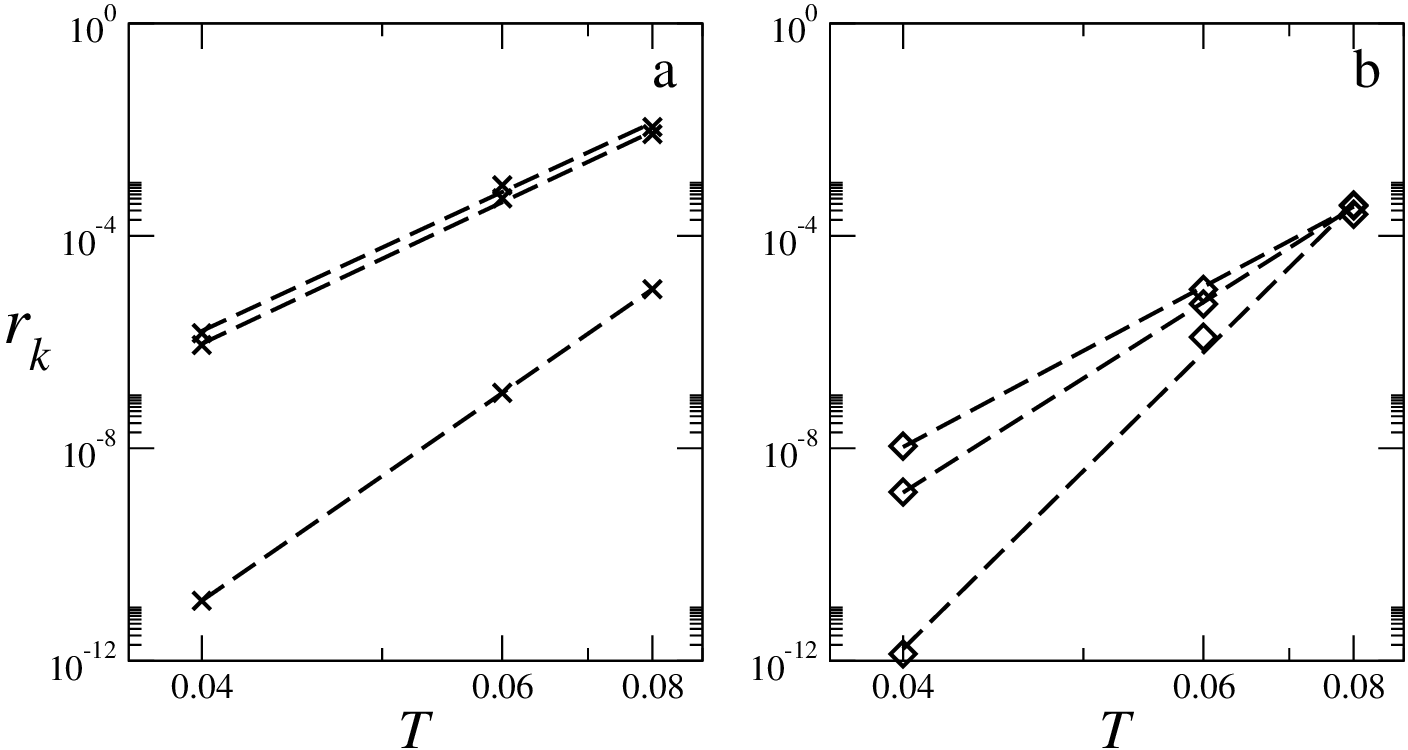}
\caption{The eigenvalues $r_k$, for $k=2,3,4$ S1 versus temperature $T$ in log--reciprocal
scale for S1 (a) and S2 (b)~. The dashed lines are exponential fits to the data.}
\label{rates}
\end{figure}
In Fig.(\ref{rates}) we show the dependence of the smallest non-zero
eigenvalues of $\tilde W$ on the temperature $T$ for sequences S1 and S2~.
In both cases we obtain evidence of an Arrhenius--like behavior  
\begin{equation}
r_k \simeq A \exp{(-B/K_B T)}.
\label{Arrlike}
\end{equation}
where $A$ and $B$ are suitable constants, which depend on the sequence and on $k$.

\begin{figure}[h]
\includegraphics[clip,width=8.7cm]{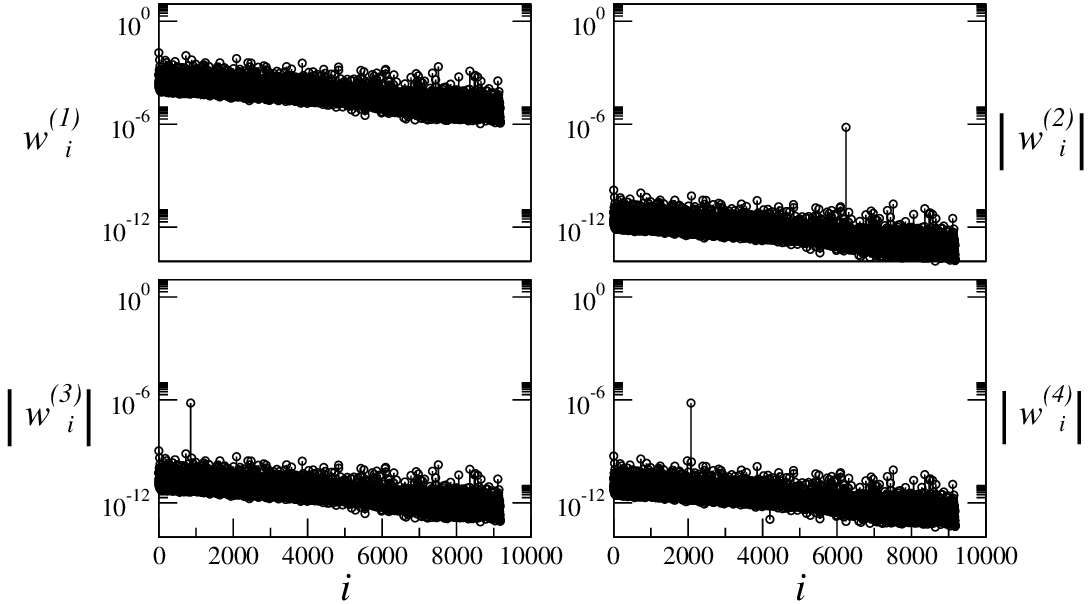}										   
\caption{Components of $w^{(1)}$ and absolute value of the components of $w^{(2)}$, $w^{(3)}$ and $w^{(4)}$ for
the slow--folder S2 at $T=0.06$.} 									   
\label{autovec01-06}
\end{figure}

\begin{figure}[h]														   
\includegraphics[clip,width=8.7cm]{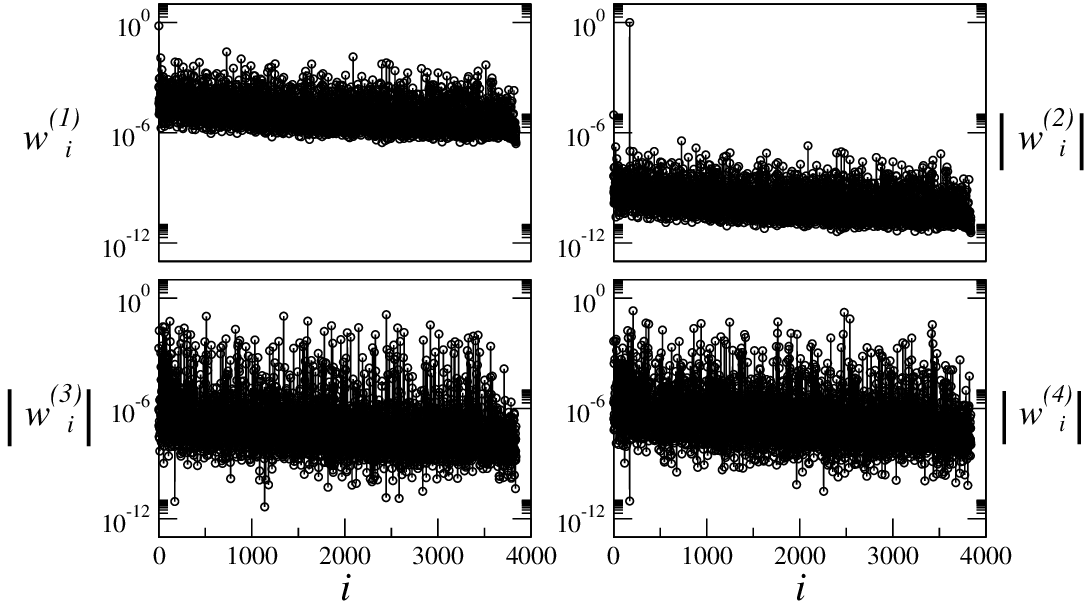}										   
\caption{Components of $w^{(1)}$ and absolute value of the components of $w^{(2)}$, $w^{(3)}$ and $w^{(4)}$ for
the fast--folder S1 at $T=0.04$.} 									   
\label{autovec81-04}														   
\end{figure}

The corresponding eigenvectors are shown in Fig.s (\ref{autovec01-06}) and
(\ref{autovec81-04}). To ease reading we report the absolute value of the components of
eigenvectors $w^{(2)}$, $w^{(3)}$ and $w^{(4)}$, since, as shown in Appendix \ref{appeA},
their components sum to 0 and therefore fluctuate wildly between positive and negative
values. The eigenvectors $w^{(1)}$ is also reported for comparison. We recall that the
eigenvector components are  ordered according to the energy of the corresponding graph node.
All the eigenvectors  of S2 are localized on some node  of the graph, as well as the
$w^{(2)}$ eigenvector of S1. The other eigenvectors of S1 are instead delocalized.

We have verified that, for the localized eigenvectors, the  energy $B$ amounts  to the
height of the lowest energy barrier separating the localized node from the graph. 
Nonetheless this energy barrier is quite high and, in this sense, the node can be viewed
as a kinetic trap of GD. We have also found that most of 
the first 50 non--zero eigenvectors of S2 are localized as those shown in 
Fig. (\ref{autovec01-06}). Accordingly, this suggest that a great deal of
nodes play the role of a kinetic trap, thus slowing-down any dynamical process on
the graph.  

\begin{figure}[h]														   
\includegraphics[clip,width=9cm]{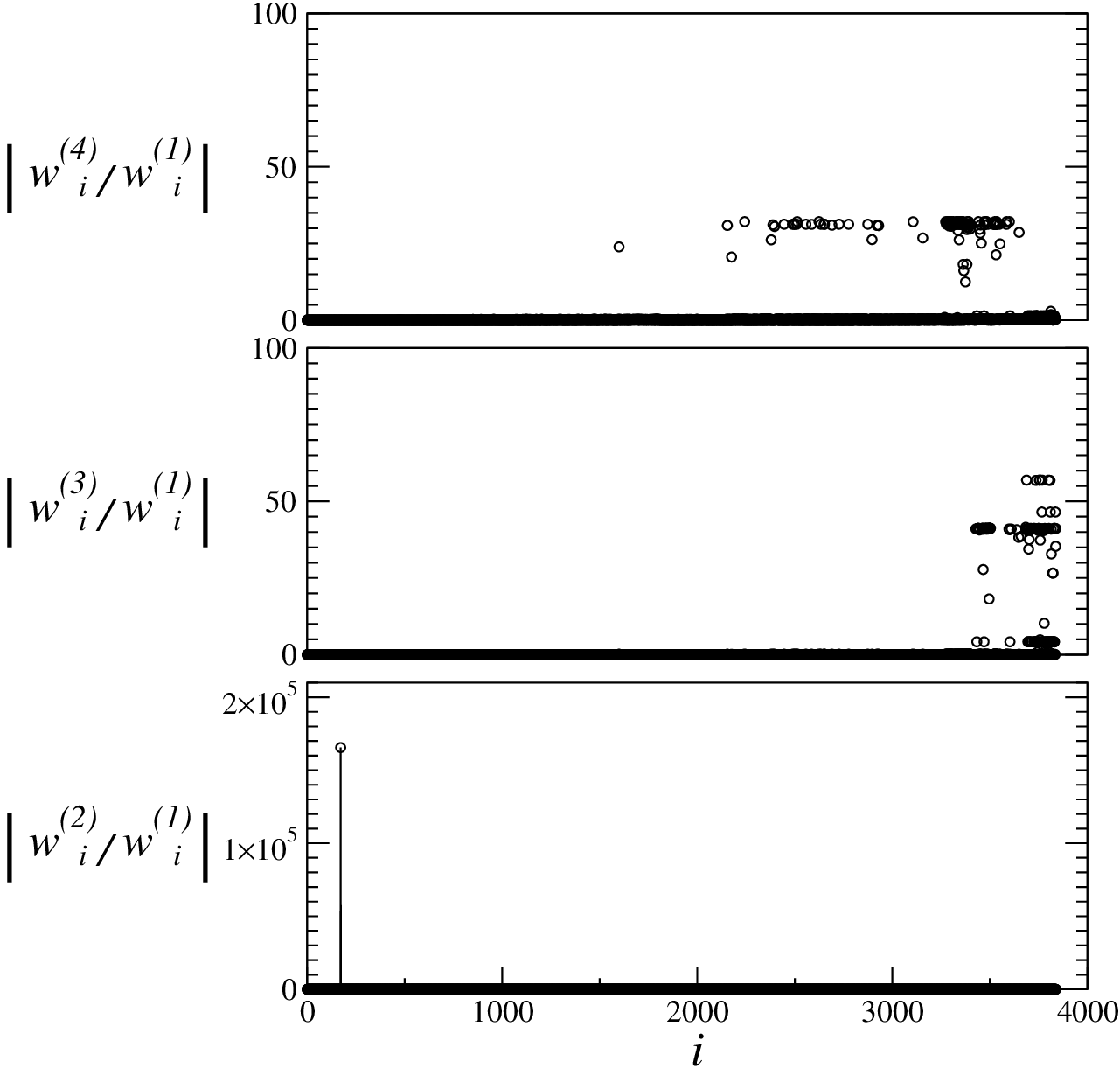}										   
\caption{Absolute value of the components of the normalized eigenvectors
$\bar{w^{(2)}}$, $\bar{w^{(3)}}$ and $\bar{w^{(4)}}$ for
the fast--folder S1 at $T=0.04$.} 									   
\label{autovec81-04weight}														   
\end{figure}

For what concerns the delocalized eigenvectors of S1, $B$ can be interpreted as
an effective energy barrier separating different subsets of nodes in the graph.
This interpretation can be supported by the following argument. Let us suppose
that a graph is subdivided into two subgraphs $\mathcal{A}$ and $\mathcal{B}$,
weakly connected between themselves. By dividing each component $w^{(2)}_i$ by
$w^{(1)}_i$ one obtains a ''normalized"  eigenvector $\bar w^{(2)}$, whose
components are split into two sets of values (see Appendix \ref{appeB} for
details)~, which identify the nodes contained into the two subgraphs. The
normalized eigenvectors $\bar{w^{(3)}}$ and $\bar{w^{(4)}}$, shown in the upper
panels of Fig.(\ref{autovec81-04weight}), exhibit a similar scenario, although
they are split into more than two sets of values. This indicates that the graph
structure of S1 is more intricated than in the example discussed in Appendix
\ref{appeB}. Nonetheless, the interpretation of $B$ as an effective barrier
height separating different regions of the graph remains valid. Notice that the
corresponding eigenvalues $r^{3}$ and $r^{4}$ are more  than four orders of
magnitude larger than $r^{2}$. This shows that the perturbative argument
presented in the Appendix provides only a qualitative approximation of what seen in 
Fig.(\ref{autovec81-04weight}).

For what concerns the folding process, we have already observed in Section \ref{sectre}
that the time scales of equilibration are orders of magnitude larger than those 
characterizing the first passage time through the native valley (i.e. the native
minimum and the connected minima). In fact, according to our definition of the folding
temperature, we can guess that over the equilibration time scale approximately $50\%$
of the time  is spent in the native  valley, despite it contains a very small fraction
of the minima (nodes) in the landscape (directed graph).  In the renormalized
representation of the directed graph we expect that the quantitative determination of
the average first passage time through the native valley (that is reduced to a single
node, as a result of the renormalization of the minima connected to the native one) is
preserved, provided the graph is renormalized for temperatures $T \leq T_f$. For
instance, we have  verified that this is the case for S1 at $T=0.04$ where the folding
time on the renormalized graph amounts to 1892 when averaged over $10^4$ paths.

\section{Conclusions and Perspectives}
\label{conclusions}

Many phenomena of biological interest are associated with equilibrium and
non--equilibrium processes in polypeptidic chains. A suitable description and
understanding of such processes is far from trivial in these complex structures. This
is the main reason why in this manuscript we have decided to consider a sufficiently
simple and widely investigated  model \cite{Still} for testing  an effective approach
to these problems. In particular, we have analyzed  two heteropolymers and one
homopolymer in order to point out differences and analogies among various typical
polypeptidic sequences. In fact, one of the heteropolymer is known to behave as a ''fast
folder", at variance with the other ones, which exhibit a much slower  relaxation
dynamics to their native states. 

As a first step we have described a strategy for reconstructing the energy landscape of
these simple chains. The search for minima and first order saddles is performed by
combining  different algorithms aiming at a sufficiently careful reconstruction of the
landscape close to the native minimum, up to values of the energy of the order of $K_B
T_f$, where $T_f$ denotes the folding temperature. Since the number of minima and
saddles increases with the energy,  the computational cost is already
quite high up to $K_B T_f$: going beyond this value is practically unfeasible. On the
other hand, performing a more extended search is not expected  to add relevant
information. In fact, we have checked that the main dynamical mechanisms associated
with the folding process and with the relaxation to equilibrium are very well
reproduced, even if most of the stationary points in the energy landscape above $K_B
T_f$ are discarded.

The following step amounts to exploit our knowledge of the relevant part of the energy
landscape for reconstructing a directed graph representation of the dynamics. Actually,
the temperature dependent molecular dynamics performed in the energy landscape can be
replaced by a Markov-chain dynamics on a directed graph. The nodes of the graph
correspond to the local minima of the energy landscape, while the first order saddles
connecting such minima are represented by the links of the graph. The strength of the
links is measured by the Langer's estimate \cite{hanggi} of the hopping rates between
connected minima. The directed nature of the graph is associated with the asymmetry due
to the different energies of the two connected minima. We have shown that, for
temperatures close or below $T_f$, MD simulations are essentially in good quantitative
agreement with GD. Anyway, in general the latter systematically underestimate the
former, mainly because it has not access to the poorly reconstructed portion of the
energy landscape above $K_B T_f$.

The main advantage of using a graph representation is that one can apply a
renormalization procedure, which keeps the topological as well as the dynamical
features of the model, while significantly reducing the dimensionality of the graph as
temperature is increased. Actually, the procedure described in Section \ref{secquat}
allows to aggregate many of the original nodes into renormalized "hubs", characterized
by a high connectivity, while a great deal of nodes exhibits a weak connectivity. This
effect is much more relevant in the  heteropolymers, than in the homopolymer, thus
indicating that the topological properties are quite different in the two cases.
Anyway, this allows us to conclude that the homopolymer, thanks to its spatially
homogeneous structure, exhibits peculiar topological properties with respect to any
disordered sequence of peptides. This notwithstanding, in the simple model considered
in this paper we have evidence that topological properties are not sufficient for
discriminating between a ''fast" and a ''slow" folder, which are known to correspond to
sequences S1 and S2, respectively. This does not exclude that topological indicators
should not be effective in more realistic models of proteins. Nonetheless, we can
affirm that dynamical properties analyzed in the MD, GD and renormalized GD provide  a
clear  identification of the ''fast folder" and are expected to be effective for the
wide class of models of polypeptides  considered in the literature. In particular, we
find that  the average first passage time from the native configuration, which provides
an estimate of the folding time, is at least two orders of magnitude smaller for the 
''fast folder" S1 than for the other sequences. By comparing this time scale with the
typical equilibration time scale associated with the inverse of the smallest eigenvalue
of the Laplacian matrix of the renormalized graph we find that they are comparable for
S0 and S2, while for S1 it is three orders of magnitude smaller. Considering that for
S1 at $T_f$ more than $50\%$ of the equilibrium measure is concentrated in the native
valley, we can conclude that the dynamics of the fast folder spends a great deal of
time exploring configurations close to its native state, even during the transient to
equilibration. This points out that the folding process can be viewed as a genuine
nonequilibrium process, since the equilibration time scales are too large to be
compared with experimental estimates of the typical folding time of a polypeptidic
chain.

We want to conclude by pointing out that the methods described in this paper can be
applied also to more realistic models of polypeptides and single--domain proteins. It
is evident that the reconstruction of the relevant portion of the energy landscape
associated with the folding and equilibration processes may yield high computational
costs. Nonetheless, the renormalization procedure can provide an effective
representation of the kinetics of these models, up to the time scales typical of the
equilibration process, which usually cannot be explored by the traditional MD
approaches.

\acknowledgments

We acknowledge CINECA in Bologna and INFM for providing us access to the
Beowulf Linux-cluster under the grant ,,Iniziativa Calcolo Parallelo''.
This work has been partially supported by the European Community via the STREP project
EMBIO (NEST contract N. 12835).


\appendix

\section{Properties of the Master Equation}
\label{appeA}
We now show some useful properties of the Master equation and of the eigenvectors of the 
Laplacian matrix.

As a preliminary observation we note that, by summing both sides of the Master equation over all 
the nodes of the graph and invoking the detailed balance condition, one can easily verify that the 
total probability is conserved:
\begin{equation}
{d(\sum_i P_i)\over dt}= 
-\sum_{i,j=1}^N P_i \Gamma_{i\rightarrow j}+\sum_{i,j}P_j\Gamma_{j\rightarrow i}=0.
\end{equation}
The normalization condition $\sum_i P_i(t)=1$ therefore holds at every $t$, which, as we will see later, 
induces some constraints on the projections of realistic probability vectors on the eigenvectors
of the Laplacian Matrix $W$.

As already mentioned, $W$ can be cast into a symmetric form trough a similarity transformation. 
It therefore admits a complete basis of orthogonal eigenvectors, each describing a different mode
of decay to equilibrium. Besides orthogonality, the eigenvectors of $W$ share the additional
property that their components are zero-sum. In fact, from the eigenvalue equation $W w^{(j)}_i =
\lambda^{(j)}w^{(j)}_i$ one gets
\begin{equation}
w^{(j)}_i={-\sum_{k=1}^N w^{(j)}_i \Gamma_{i,k}+\sum_{k=1}^N w^{(j)}_k\Gamma_{k,i} \over
\lambda^{(j)}}.
\end{equation}
Summing over $i$ each members and once again invoking detailed balance, leads to
\begin{equation}
\sum_{i=1}^N w^{(j)}_i=0.
\end{equation}

The only eigenvector that defies this demonstration is $w^{(1)}$ the null eigenvector defined in Equation 
(\ref{stationary}), which actually has positive components and can be normalized to unity. Using this
normalization one can write:
\begin{equation}
\sum_{i=1}^N w^{(j)}_i=\delta_{1,j}.
\end{equation}
Actually the fact that eigenvector have zero sum is just a consequence of the fact that the 
master equation conserves probability. Indeed, since eigenvectors form a complete basis, each
probability distribution of initial conditions on the graph  $P(0)$ can be expressed as  
$P(0)=\sum_{j=1}^N \alpha_j w^{(j)}$. It will then evolve in time according to
\begin{equation}
P(t)=\sum_{j=1}^N \alpha_j w^{(j)} \exp^{-r_j t}.
\end{equation}

Summing over the components of $P(t)$ 
\begin{equation}
\sum_{i=1}^N P_i(t)= \sum_{j=1}^N\alpha_j \;\delta_{1,j} \exp^{-r_j t} =\alpha_0.
\end{equation}
Hence, in order to have $\sum_{i=1}^N P_i(t)=1$,  $\alpha_1$ must necessarily be 1.

\section{Spectral clustering} 
\label{appeB}
We now justify the use of reweighted components as an effective  tool to
uncover the inherent structure of eigenvectors. We define reweighted components of  a vector $v$ on a graph
as the ratio site--to--site of the vector component to the local value  of the stationary probability:
$\bar{v_i}=v_i/w_i^{(0)}$. We will here extend an argument originally  proposed for discrete graphs
\cite{ding} to weighted ones and show that, when a graph is divided into two weakly connected subgraphs $\mathcal{A}$
and $\mathcal{B}$, the reweighted components  of the first nonzero eigenvector
assume only two possible values one for $\mathcal{A}$ and one for $\mathcal{B}$.


The Laplacian Matrix $W$ can be written as the sum of two matrices: $W=D-\Gamma^T$, where
$\Gamma$ is the transition rate matrix and 
$D$ a diagonal matrix $D_{i,j}=\delta_{i,j}\sum_{k=1}^N \Gamma_{i,k}$.

First of all, we consider the case in which the graph is composed of two 
disconnected subsets of nodes $\mathcal{A}$ and $\mathcal{B}$. In this case the Laplacian matrix will be referred to 
as $W^0$. Since $\Gamma_{i,j}=0$ for each $i \in \mathcal{A}$ and $j \in \mathcal{B}$, $W^0$ can be written as
 \begin{equation}
W^0=\left(\begin{array}{cc}
D_{\mathcal{AA}} - \Gamma_{\mathcal{AA}}^T & 0 \\
0 & D_{\mathcal{BB}} - \Gamma_{\mathcal{BB}}^T  
\end{array}\right)
\end{equation}
Let now use the null eigenvector $w^{(1)}$ of $W^0$ to construct two vectors $w^\mathcal{A}$ e $w^\mathcal{B}$ as follows:
\begin{equation}
w_i^\mathcal{A}=\left\{ \begin{array} {lc}
w_i^{(1)}  & i\in \mathcal{A}\\
0        & i\in \mathcal{B}
\end{array}\right.,\;\;
w_i^\mathcal{B}=\left\{ \begin{array} {lc}
0        & i\in \mathcal{A}\\
w_i^{(1)}  & i\in \mathcal{B}
\end{array}\right.
\end{equation}
It is easy to show that both $w^\mathcal{A}$ and $w^\mathcal{B}$ are eigenvectors of $W^0$ with a null
eigenvector, and the same holds for any linear combination $v=a w^\mathcal{A}+ b w^\mathcal{B}$.
More generally when a graph is composed of $n$ disconnected subgraphs the kernel
of its Laplacian matrix is has dimension $n$.

Let's now suppose that the two subgraphs $\mathcal{A}$ and $\mathcal{B}$ are not properly disconnected
but do share a small number of connections. In this case
the Laplacian matrix has the form $W=W^0+W^1$, with
 \begin{equation}
W^1=\left(\begin{array}{cc}
D_{\mathcal{AB}} & -\Gamma_{\mathcal{AB}}^T  \\
 -\Gamma_{\mathcal{BA}}^T& D_{\mathcal{BA}}   
\end{array}\right).
\end{equation}
where $\Gamma_{\mathcal{AB}}$ and $\Gamma_{\mathcal{BA}}$ carry the information about connections between $\mathcal{A}$ 
and $B$, while $D_{\mathcal{AB}}$ and $D_{\mathcal{BA}}$ carry the information on the effect these connections
have on the diagonal of the Laplacian matrix.

We now look for the eigenvectors of $W$ among vectors of the form  $v = a w^\mathcal{A} + b w^\mathcal{B}$:
$$
W v = (W^0 + W^1) (a w^\mathcal{A}+ b w^\mathcal{B})=
$$
\begin{equation}
= W^1 (a w^\mathcal{A} + b w^\mathcal{B}). 
\label{equivalenceWW1}
\end{equation}
In other words if any eigenvector of $W$ exists that is a linear combination of $w^\mathcal{A}$ and $w^\mathcal{B}$
it is also an eigenvector of $W^1$. We will therefore look for the eigenvectors of this last matrix 
having the desired form  $a w^\mathcal{A} + b w^\mathcal{B}$.

For sufficiently small $W^1$ the vector $W^1 (a w^\mathcal{A} + b w^\mathcal{B})$ can be approximately written as a linear 
combination of $w^\mathcal{A}$ and $w^\mathcal{B}$. To this purpose we introduce the projector on the space of the linear 
combinations of $w^\mathcal{A}$ and $w^\mathcal{B}$
\begin{equation}
\Pi_{\mathcal{A}\mathcal{B}}=(w^\mathcal{A},w^\mathcal{B})
\end{equation}
where $(w^\mathcal{A},w^\mathcal{B})$ is a $N\times2$ matrix whose columns are the two column vectors $w^\mathcal{A}$ and $w^\mathcal{B}$.
The vector $W^1 (a w^\mathcal{A} + b w^\mathcal{B})$ can now be written as:
$$
W^1 (a w^\mathcal{A} + b w^\mathcal{B}) \simeq \Pi_{\mathcal{AB}} W^1 (a w^\mathcal{A} + b w^\mathcal{B})=
$$
\begin{equation}
=
\tilde{W^1} \left(\begin{array}{c}a\\b\end{array}\right) (w^\mathcal{A},w^\mathcal{B})
\end{equation}
where we have introduced $\tilde{W^1}=\Pi_{\mathcal{AB}}^T W^1 \Pi_{\mathcal{AB}}$, 
a $2\times 2$ matrix that reproduces the effect of $W^1$ 
in the subspace of the linear combinations of $w^\mathcal{A}$ and $w^\mathcal{B}$.
After some algebra $\tilde{W^1}$ reads
\begin{equation}
\tilde{W^1} = {\small
\left(\begin{array}{cc}
 \sum_{i \in \mathcal{A}, j \in \mathcal{B}} {w^{(1)}_i}^2 \Gamma_{i,j}       &   -\sum_{i \in \mathcal{A}, j \in \mathcal{B}} w^{(1)}_iw^{(1)}_j \Gamma_{j,i}   \\
+\sum_{i \in \mathcal{B}, j \in \mathcal{A}} w^{(1)}_i w^{(1)}_j \Gamma_{j,i} &    \sum_{i \in \mathcal{B}, j \in \mathcal{A}} {w^{(1)}_i}^2 \Gamma_{i,j} 
\end{array}\right)}.
\end{equation}
Since $w^{(1)}$ satisfies the detailed balance, and existing a connection from
$\mathcal{B}$ to $\mathcal{A}$ for each connection from $\mathcal{A}$ to $\mathcal{B}$, one has:
\begin{equation}
\sum_{j \in \mathcal{B}} w^{(1)}_i \Gamma_{i,j} = 
\sum_{j \in \mathcal{B}} w^{(1)}_j \Gamma_{j,i}
\;\;\;\;\;\;\forall i \in \mathcal{A}.
\end{equation}
Analogously:
\begin{equation}
\sum_{j \in \mathcal{A}} w^{(1)}_i \Gamma_{i,j} = 
\sum_{j \in \mathcal{A}} w^{(1)}_j \Gamma_{j,i},
\;\;\;\;\;\;\forall i \in \mathcal{B}.
\end{equation}
We can therefore define two quantities 
$\alpha=\sum_{i \in \mathcal{A}, j \in \mathcal{B}} w^{(1)}_i w^{(1)}_j \Gamma_{j,i}$ and 
$\beta=\sum_{i \in \mathcal{B}, j \in \mathcal{A}} w^{(1)}_i w^{(1)}_j \Gamma_{j,i}$ which cast $\tilde{W^1}$  in a particularly simple form
\begin{equation}
\tilde{W^1} = 
\left(\begin{array}{cc}
      \alpha &    -\alpha  \\
    - \beta  &    \beta
\end{array}\right)
\end{equation}
It is important to notice that, if the two subgraphs $\mathcal{A}$ are weekly connected $\alpha$ and $\beta$
will be relatively small, since there are few connections such that $\Gamma_{j,i}>0$ for $i\in \mathcal{A}$ 
and $j\in \mathcal{B}$.

The two eigenvalues of $\tilde{W^1}$ are 0 and $\alpha+\beta$, referring respectively to
the eigenvectors 
\begin{equation}
\left(\begin{array}{c}1\\1\end{array}\right)  
\;\;\;\;\;\mathrm{and}\;\;\;\;\; 
\left(\begin{array}{c}  \alpha \\-\beta\end{array}\right).
\end{equation}
It can finally be shown that backprojecting with $\Pi_{\mathcal{AB}}^T$ this two eigenvectors of $\tilde{W^1}$ to the 
entire $N$-dimensional space one obtains two eigenvectors of $W^1$ characterized by the same eigenvalues. 
According to Equation \ref{equivalenceWW1} these are also eigenvectors of $W$. 
More precisely by this procedure one obtains
\begin{itemize}
\item $w^\mathcal{A}+w^\mathcal{B}$ which obviously coincides with $w^{(1)}$ and is associated to the null
eigenvalue also according to the perturbative calculation
\item  $u=\alpha w^\mathcal{A} - \beta w^\mathcal{B}$, associated to the eigenvalue $\alpha+\beta$ which is a small 
number and  will therefore lay in the low end of the spectrum of $W$.
\end{itemize}
It is now straightforward to verify that the reweighted coordinates of $u$ get the values $\alpha$ 
for nodes belonging to $\mathcal{A}$ and $-\beta$ for nodes belonging to $\mathcal{B}$. In this sense the analysis of the 
reweighted coordinates of the eigenvectors of $W$ can be employed as a spectral method for the identification
of clusters, portion of the graph characterized by a high degree of internal connectivity and a small
number of connections with the rest of the graph.


%


\end{document}